\documentclass[a4paper,12pt]{article}
\usepackage{graphicx}
\usepackage{bm}
\def\be{\begin{equation}}
\def\ee{\end{equation}}
\def\bea{\begin{eqnarray}}
\def\eea{\end{eqnarray}}
\usepackage{latexsym}
\usepackage{dcolumn}
\usepackage{epsfig,amssymb,euscript}
\usepackage{amsmath}
\usepackage{array,calc,epsfig}

\usepackage[noadjust]{cite}

\usepackage{chngcntr}

\usepackage{color}

\renewcommand{\to}{\rightarrow}
\renewcommand{\l}{\lambda}

\def\be{\begin{equation}}
\def\ee{\end{equation}}
\def\ba{\begin{eqnarray}}
\def\ea{\end{eqnarray}}
\def\nb{\nonumber}

\def\a{\alpha}
\def\b{\beta}

\def\ff{\phi}

\def\g{\gamma}
\def\G{\Gamma}

\def\l{\lambda}

\def\n{\nu}
\def\o{\omega}
\def\O{\Omega}
\def\r{\rho}
\def\s{\sigma}

\def\q{\quad}

\newcommand{\pr}[1]{\left(#1\right)}
\newcommand{\pq}[1]{\left[#1\right]}

\counterwithin*{equation}{section}

\begin{document}
\baselineskip=15.5pt
\pagestyle{plain}
\setcounter{page}{1}
\newfont{\namefont}{cmr10}
\newfont{\addfont}{cmti7 scaled 1440}
\newfont{\boldmathfont}{cmbx10}
\newfont{\headfontb}{cmbx10 scaled 1728}
\renewcommand{\theequation}{{\rm\thesection.\arabic{equation}}}
\renewcommand{\thefootnote}{\arabic{footnote}}

\vspace{1cm}
\begin{titlepage}
\vskip 2cm
\begin{center}
{\Large{\bf Dark Holograms and Gravitational Waves}}
\end{center}

\vskip 10pt
\begin{center}
Francesco Bigazzi$^{a}$, Alessio Caddeo$^{a,b}$, Aldo L. Cotrone$^{a,b}$, Angel Paredes$^{c}$
\end{center}
\vskip 10pt
\begin{center}
\vspace{0.2cm}
\textit {$^a$ INFN, Sezione di Firenze; Via G. Sansone 1; I-50019 Sesto Fiorentino (Firenze), Italy.
}\\
\textit{$^b$ Dipartimento di Fisica e Astronomia, Universit\'a di Firenze; Via G. Sansone 1;\\ I-50019 Sesto Fiorentino (Firenze), Italy.
}\\
\textit{$^c$ Departamento  de  Fisica  Aplicada,  Universidade  de  Vigo,  As  Lagoas  s/n,  Ourense,  ES-32004  Spain.}\\
\vskip 20pt
{\small{
bigazzi@fi.infn.it, alessio.caddeo@unifi.it, cotrone@fi.infn.it, angel.paredes@uvigo.es
}

}

\end{center}

\vspace{25pt}

\begin{center}
 \textbf{Abstract}
\end{center}

\noindent 
Spectra of stochastic gravitational waves (GW) generated in cosmological first-order phase transitions are computed within strongly correlated theories with a dual holographic description.
The theories are mostly used as models of dark sectors.
In particular, we consider the so-called Witten-Sakai-Sugimoto model, a $SU(N)$ gauge theory coupled to different matter fields in both the fundamental and the adjoint representations. The model has a well-known top-down holographic dual description which allows us to perform reliable calculations in the strongly coupled regime. We consider the GW spectra from bubble collisions and sound waves arising from two different kinds of first-order phase transitions: a confinement/deconfinement one and a chiral symmetry breaking/restoration one. Depending on the model parameters, we find that the GW spectra may fall within the sensibility region of ground-based and space-based interferometers, as well as of Pulsar Timing Arrays. 
In the latter case, the signal could be compatible with the recent potential observation by NANOGrav.
When the two phase transitions happen at different critical temperatures, characteristic spectra with double frequency peaks show up. Moreover, in this case we explicitly show how to correct the redshift factors appearing in the formulae for the GW power spectra to account for the fact that adiabatic expansion from the first transition to the present times cannot be assumed anymore.

\end{titlepage}
\newpage
\tableofcontents

\section{Introduction}

The measurement of the first direct gravitational wave (GW) signal by LIGO in 2015 \cite{Abbott:2016blz} has started a new era in observational astrophysics.
Not only the observation of black hole and neutron star mergers are tremendously important discoveries, but current and future experiments are now expected to be able to measure GW signals from several different sources.
This promises to give experimental access to physics which would be challenging to investigate with other types of observations.
Not surprisingly, there are currently several experiments in the developing phase, which will considerably extend the accessible GW frequency and sensitivity ranges in the near future.   
In this situation, it is of clear interest to study possible sources of GWs which could be detected in these facilities.

In this paper, we consider stochastic GW spectra produced in first-order cosmological phase transitions.
The generation of GWs, in this case, is determined by the dynamics of bubbles of true vacuum nucleated in the metastable phase once the temperature of the Universe descends below the phase transition temperature \cite{Coleman:1977py, Callan:1977pt, Coleman:1980aw,Linde:1980tt,Linde:1981zj}.
The bubbles can generate GWs either by their collisions or by their interaction with the plasma medium, through sound waves or turbulence.
We refer to \cite{Maggiore:2018sht,Caprini:2015zlo,Caprini:2019egz,Hindmarsh:2020hop} for reviews.

It is a  challenging task to connect the qualitative picture of the bubble dynamics to solid predictions for the power spectra of GWs that can be observed in experimental devices. Luckily, there are general formulae in the literature that estimate the GW spectra once some parameters characterizing the phase transition are known. These parameters depend on the details of the microscopic model describing the transition. The evaluation of the parameters and the formulae for the spectra typically rely on a series of controlled and less controlled approximations. It is a crucial  goal to reduce to zero the number of uncontrolled approximations such that the theoretical predictions can be reliably tested in experiments.

This paper makes a step in this direction for cosmological transitions in sectors described by strongly-coupled Yang-Mills or QCD-like theories.
The latter appear in many dark matter models (see, e.g., \cite{Cline:2012bz,Kribs:2016cew,Profumo:2019ujg}). We consider scenarios where the dark matter is constituted e.g.~by dark glueballs, pions or baryons. 

Whenever the theory is confining, one expects a confinement/deconfinement transition as the Universe cools below the theory's dynamical scale. If the transition is first order, it may generate GWs, which are the study objects of this paper.

When the gauge theory includes (approximately) massless quarks, the strongly-coupled dynamics is such that the (approximate) chiral symmetry is broken at a scale that might or might not coincide with the gauge theory's dynamical scale. We consider both the case in which the confinement phase transition implies the chiral symmetry phase transition and the case in which it does not. The first case also includes the Peccei-Quinn transition in the simplest composite axion model with hidden gauge group \cite{Kim:1979if,Kaplan:1985dv,Choi:1985cb}. The second case includes the Peccei-Quinn first-order phase transition of the recently proposed  holographic axion model \cite{Bigazzi:2019eks,Bigazzi:2019hav}, where the axion appears as a pseudo-Nambu-Goldstone boson associated with the chiral symmetry breaking of an extra pair of quark/antiquark fields.

To be more specific, we consider theories where the rank of the gauge group is sufficiently large such that the planar approximation is reliable. 
In this case, a class of very interesting models are the ones admitting a holographic description.\footnote{It is a widespread belief that every gauge theory in the planar limit admits a perturbative string theory description. The latter can or cannot have a low-energy limit where classical gravity is reliable, depending on the theory's details. For example, pure Yang-Mills theory does not have a classical gravitational description. Nevertheless, there are infinite classes of theories,  which we call \emph{holographic}, admitting such a description.} 
As a prototype,
we consider the top-down theory that, in the deep IR, better resembles planar Yang-Mills, or planar QCD if we consider the flavored version, known as the Witten-Sakai-Sugimoto (WSS) model \cite{Witten:1998zw,Sakai:2004cn}.
In the case of a YM or QCD-like dark sector outside the regime where the holographic description is completely reliable, the latter can be employed as an effective approach to the strong dynamics of the theory.

The WSS model has been widely used to study various aspects of QCD at low energy, with notable success. In the present context, the model is interesting because it features two first-order phase transitions, the first one associated with confinement/deconfinement and the second one associated with chiral symmetry breaking/restoration.

In most of the cases that we investigate, we employ the WSS model not as a proxy for QCD but as a model for a dark sector.
Being a so-called top-down model, the WSS has the advantage that computations performed in the planar limit at strong coupling are reliable, in the sense that there is a precise control on the validity regime of the various approximations, something which usually does not occur in effective phenomenological models or bottom-up holographic theories. In fact, this property eliminates one of the sources of uncertainty in the calculation of the parameters for the GWs spectra when dealing with strongly-coupled theories, and it constitutes the main motivation for this paper.

In a previous work \cite{Bigazzi:2020phm}, we have addressed the problem of the nucleation of bubbles of true vacuum associated with both the confinement/deconfinement phase transition and the chiral symmetry breaking/restoration phase transition in the WSS model.\footnote{In \cite{Bigazzi:2020phm}, the Randall-Sundrum scenario has been briefly discussed as well. In that context, we were able to compute the derivative term in the effective bounce action in the high temperature regime. That term was missing in previous literature on the subject.}  

In the present work, we use those results to compute the stochastic GW spectra, due to bubble collisions and sound waves, in several beyond Standard Model scenarios featuring the WSS model. As we will see, the main conclusion of our analysis is that there is a large window of the WSS parameter space where the GW signals may be accessible in near-future experiments.
Moreover, the model allows for the generation of GWs compatible with the possible observation recently reported by NANOGrav \cite{nanograv}.

The paper is organized as follows.
In section \ref{basicsecWSSreview}, we introduce the WSS model and summarize the steps of the analysis needed to find the GW spectra.

In section \ref{darkscenarios}, we consider three different dark matter scenarios. These are cases where the chiral symmetry transition, if present, is implied by the confinement one.
In subsection \ref{secresultsconfinement}, we discuss the results for the GW spectra. 
Figure \ref{figtutti} encodes in a global view some benchmark results of the investigation.

In section \ref{sec:chisb}, we consider two scenarios where GWs come from the chiral symmetry breaking/restoration phase transition. In one of them, the chiral transition is followed by a separated confinement/deconfinement one. We thus investigate the fascinating possibility of detecting a GW spectrum with two peaks. In this case, moreover, we outline the fact that the usual assumption of adiabatic expansion of the Universe from the first phase transition to present times cannot be used anymore: the presence of a second phase transition requires a refinement of the usual redshift factors in the formulae for the GW spectra. The results for the GW spectra are reported in subsection \ref{secresultschiral} and in figure \ref{experimentchirallittle}.

We will conclude with a summary and some observations in section \ref{secconclusions}.

We collect all the holography-related details of the WSS model in appendix \ref{moresecWSSreview}. Appendix \ref{appformule} provides an overview of all of the relevant formulae used to obtain the GW spectra. In particular, in \ref{sec:redshift}, we discuss how the occurrence of two separated phase transitions affects the quantities that determine the GW spectra, providing explicit formulae for the modified redshift factors advocated in section \ref{sec:chisb}.
Finally, in appendix \ref{sec:holobubbles}, we review the results of \cite{Bigazzi:2020phm} that are useful for the present paper and provide approximate analytical expressions of the relevant GW parameters for the confinement/deconfinement transition in the small temperature regime.


\section{The WSS model and its embedding in cosmology}
\label{basicsecWSSreview}
In this section, we describe the features of the Witten-Sakai-Sugimoto model that are needed in order to understand the calculation of the GW spectra.
More details on the model and on the bubble configurations nucleated in the phase transitions are reported in appendices \ref{moresecWSSreview} and \ref{sec:holobubbles}. 

The WSS model is a (3+1)-dimensional non-supersymmetric Yang-Mills theory with gauge group $SU(N)$ coupled to a tower of adjoint Kaluza-Klein (KK) fields and to $N_f$ fundamental flavors (quarks) \cite{Witten:1998zw,Sakai:2004cn} (see also \cite{Rebhan:2014rxa} for a review). The model possesses five independent parameters. Two of them are actually dimensional quantities: $M_{KK}$, which represents the dynamically generated scale providing the mass of the first glueball and that of the first KK field, and $L$ which gives the scale of chiral symmetry breaking $f_{\chi, L}$, as we will discuss in a moment.
The other three, dimensionless parameters are given by $N$, $N_f$, and the 't Hooft coupling $\l$ at the scale $M_{KK}$. We will consider the regime
\be
\label{regimeofstudy}
N\gg1 \ , \q \q \q   \lambda \gg 1 \ , \q \q \q \frac{N_f}{N} \ll 1   \ .
\ee
The properties of the model at low energies are very similar to the real-world QCD ones since they include confinement, mass gap, and chiral symmetry breaking. We can actually write more precisely the last condition in (\ref{regimeofstudy}) as (see e.g. \cite{Bigazzi:2014qsa})
\be
\label{epsilonf}
\epsilon_f \equiv \frac{1}{12 \pi^3}\lambda^2 \frac{N_f}{N} \ll 1 \ ,
\ee
which holds in the confined regime.\footnote{In the deconfined phase, the condition reads $\epsilon_{f,T} \equiv \l^2 N_f T/ (6 \pi^2 N M_{KK}) \ll 1$ \cite{Bigazzi:2014qsa}.}

One of the main motivations for studying the model in this paper is that it exhibits two first-order phase transitions. The first one separates the low temperature confined phase of the theory from the high temperature deconfined one. The critical temperature for the transition is  \cite{Aharony:2006da}
\be
\label{confinementcriticaltemperature}
T_c = \frac{M_{KK}}{2 \pi} \ .
\ee

The second first-order phase transition separates the chirally symmetric phase from the phase where chiral symmetry is broken \cite{Aharony:2006da}. 
In the general case, $L$ is a free parameter of the model that can be used to separate the confinement scale from the chiral symmetry breaking one.
When $L> 0.97 M_{KK}^{-1}$ the confinement/deconfinement transition implies the chiral symmetry breaking/restoration one. In contrast, when $L< 0.97 M_{KK}^{-1}$, the two transitions are independent, with the chiral symmetry breaking/restoration one occurring at the temperature
\be
\label{chiralcriticaltemperature}
T_c ^\chi \approx \frac{0.1538}{L} \ .
\ee

The parameter $L$ has the maximal value $L = \pi M_{KK}^{-1}$, when the scale of chiral symmetry breaking reads 
\be
\label{decayconstantpion}
 f_\chi ^2 = \frac{\lambda N}{108 \pi^4} M_{KK} ^2 \ .
\ee
In the opposite limit $L \ll \pi M_{KK}$, we have
\cite{Aharony:2007uu, Bigazzi:2019eks}\footnote{Note that in this paper a different convention on the coupling w.r.t. \cite{Bigazzi:2019eks} is used: $\lambda_{here}=2\lambda_{there}$.} 
\be
\label{fchil}
f_{\chi, L}^2 \simeq 0.1534 \frac{\lambda N}{32 \pi^3} \frac{1}{M_{KK} L^3}\ .
\ee
So far, we have been assuming that all the $N_f$ quarks condense at the same scale, dictated by the same value of $L$. But of course we can actually have several classes of quarks with different values of $L$.

To summarize, the phase diagram of the model is the following:
\begin{itemize}
\item  If $T<\frac{M_{KK}}{2\pi}$, the theory is confining and chiral symmetry is broken;
\item If $T>\frac{M_{KK}}{2\pi}$, the theory is deconfined and:
\begin{itemize}
\item if $T<\frac{0.1538}{L}$, chiral symmetry is broken;
\item if $T>\frac{0.1538}{L}$, chiral symmetry is preserved.
\end{itemize}
\end{itemize}

\subsection{Cosmological WSS phase transitions}
\label{seccosmo}

In this subsection, we describe the general framework needed to calculate the GW spectra, also fixing our notation. 
We leave most of the technical details, which are quite standard, to appendix \ref{appformule}, for the benefit of the reader who is not familiar with this type of computations.

We will consider a cosmological setting where the Universe starts at some high temperature, in which the WSS is in the deconfined phase, and then cools down. Depending on the scenario that we consider, the WSS sector will undergo one or two first-order phase transitions. They are triggered by the nucleation of bubbles of \emph{true} vacuum (confined phase or chirally broken phase, depending on the transition) in the plasma, which is in the metastable \emph{false} vacuum (deconfined or chirally symmetric). These bubbles will expand and eventually fill all the Universe, leaving it in the true vacuum state. The \emph{percolation} temperature $T_p$ is defined as the temperature of the Universe when this process completes. We will compute it case by case, using the formulae discussed in appendix \ref{relevanttemperatures}.

The cosmological evolution of the Universe is described, as usual, by the Friedmann-Lemaitre-Robertson-Walker (FLRW) metric\footnote{As we discuss in appendix \ref{moresecWSSreview}, the WSS model is based on string-theory, where extra dimensions are involved. The cosmic scale factor is meant to be present just in front of the spatial three-dimensional space.}
\be
ds^2 = -dt^2 + R(t)^2 dx^i dx^i  \ , 
\ee
where $R(t)$ is the cosmic scale which defines the Hubble scale $H(t) = \dot{R}(t)/R(t)$. The latter is determined by the total energy density through the Friedmann equation
\be
H^2 = \frac{\rho}{3 M_{Pl}^2}\ ,
\ee
with $M_{Pl} \approx 2.4\cdot 10^{18}$ GeV. The energy density $\r$ takes contributions from the standard model and from the dark sector. 

In the sector described by the WSS model, the energy density in the deconfined and in the confined phase at order ${\cal O}(N^2)$ reads, respectively,
\begin{subequations}
\label{energydarkgauge}
\ba
\rho_{rad,glue} &=& 5\frac{2^6 \pi^4 }{3^7} \l N^2  \frac{T^6}{M_{KK}^2}   \  . \\
\rho_{conf, glue} = - \r_{0,glue} &=& - \frac{1}{3^7 \pi^2} \l N^2 M_{KK}^4  \ .
\ea
\end{subequations}
In the limit (\ref{epsilonf}), the contribution of $N_f$ quarks to the energy density in the high-temperature regime and in the low-temperature one at order ${\cal O}(N_f N)$, in the case $L = \pi M_{KK} ^{-1}$, read (see e.g. \cite{Ballon-Bayona:2013cta, Bigazzi:2014qsa}) 
\begin{subequations}
\label{energyflavors}
\ba
\label{energyflavorconfinedantipodal}
\rho_{rad,\chi} &=& \frac{2^6 \pi^2}{7\cdot 3^7} \lambda^3 N_f N \frac{T^7}{M_{KK}^3}\ , \\
\r_{conf, \chi} =
- \rho_{0,\chi} 
&=& -\frac{1}{7\cdot 3^7 \pi^{7/2}\Gamma \pr{-\frac23 } \Gamma \pr{\frac16 }} \lambda^3 N_f N M_{KK}^4 \ .
\label{energyflavorconfinedantipodal2}
\ea
\end{subequations}

As discussed above, when $L \ll \pi M_{KK} ^{-1}$, there is an intermediate phase where the gauge theory is deconfined and the quarks are condensed. In this case, the energy density is not known analytically. However, it can be computed numerically starting from the energy density of the chirally-unbroken configurations. In particular, it reads,
\begin{subequations}
\label{energydeconfcondensed}
\be
\label{energydeconfcondensed1}
\rho_{b,\chi} = \rho_{rad,\chi} + (1 - T \partial_T )(T P \Delta \tilde S)\ ,
\ee
where $TP \Delta \tilde S$ gives the difference of free energies of the flavors in the broken and unbroken phases, with
\be
T P = \frac{2^3 \pi^2}{3^8} \lambda^3 N_f N \frac{T^7}{M_{KK}^3}\ ,
\ee
\end{subequations}
and $\tilde S$ defined as in (\ref{factors_action}).
Using the fact that the energy is the derivative of the free energy w.r.t. the temperature, the second term on the r.h.s.~of  (\ref{energydeconfcondensed1}) is the difference of the energies in the two phases, so that adding the known contribution of the unbroken phase, one is left with that of the broken phase.   
As we will comment on in section \ref{secholoaxion1}, the energy density of condensed quarks with  $L = \pi M_{KK} ^{-1}$ in the confined phase will always be subleading and can be neglected.

From (\ref{energydarkgauge}) and (\ref{energyflavors}), we see that the confined phase of the WSS model carries a temperature-independent contribution to the energy, which would act as a cosmological constant after the phase transition. Since the measured cosmological constant almost vanishes, the zero-point energy has to be shifted accordingly. As a result, the energy density in the deconfined and chirally symmetric phase reads\footnote{In the most general case, we have quarks of both $L = \pi M_{KK} ^{-1}$ and $L \ll \pi M_{KK} ^{-1}$ kind.
Hence, the contribution $\rho_{0,\chi}$ is not simply given by (\ref{energyflavorconfinedantipodal2}), because the latter holds only for the $L = \pi M_{KK} ^{-1}$. The $L \ll \pi M_{KK} ^{-1}$ contribution is suppressed by a $M_{KK}/f_{\chi,L}$ factor and can be usually neglected.}
\be
\label{energytot}
\r ^{deconf} _{unbroken} = \rho_{rad,glue} + \rho_{rad,SM} + \rho_{rad,\chi} + \rho_{0, glue} + \rho_{0,\chi} \ ,
\ee
where 
\be
\rho_{rad,SM} = \frac{\pi^2}{30}g_*^{SM} (T) \frac{T^4}{\xi^4}
\ee
is the Standard Model contribution, given by the temperature-dependent number of relativistic degrees of freedom $g_*^{SM}$. The factor 
\be
\xi \equiv \frac{T}{T_V} \ ,
\ee
is defined as the ratio between the temperature $T$ of the dark sector  and that of the Standard Model $T_V$. As we will see, $\xi$ can (and in some cases must) be different from 1. 

The energy density in the deconfined and chirally broken phase reads
\be
\label{energytotdeconfinedcbroken}
\rho ^{deconf}_{broken} =  \rho_{rad,glue} + \rho_{rad,SM} + \rho_{b,\chi} + \rho_{0,glue} \ ,
\ee
whereas in the confined and chirally broken phase it is
\be
\label{energytotconfinedcbroken}
\r  ^{conf} _{broken} = \frac{\pi^2 }{30 } \left( g_*^{SM}(T) \frac{T^4}{\xi^4} +  g_* (T) T^4 \right) \ ,
\ee
where $g_* (T)$ 
accounts for possible contributions of
relativistic particles from the dark sector. 

We will investigate several scenarios where (\ref{energytot}), (\ref{energytotdeconfinedcbroken}) and (\ref{energytotconfinedcbroken}) will be used.  The cases will differ for the values of the parameters $N_f$, $N$, $\l$, and the number of degrees of freedom involved.

Away from the phase transitions, the universe evolves adiabatically, i.e. according to the conservation of the entropy
\be
S \sim R^3 g_* ^S (T) T^3 \ ,
\ee
where, in general, $g_*^S(T)\neq g_*(T)$, see appendix \ref{sec:redshift}.
During the phase transition, an amount of energy is released and the plasma gets heated up. The temperature $T_R$ of the plasma at the end of the transition is called \emph{reheating} temperature and is found via the conservation of energy. This point will play an important role in section \ref{sec:chisb}, where we will consider the case in which the universe undergoes two first-order phase transitions.
As we will see, the presence of the second phase transition modifies the redshift of the GW signal compared to the adiabatic evolution one, usually assumed to be valid after the single phase transition.

As we detail in appendix \ref{appformule}, the efficiency of the phase transition depends on the ratio $\G/H^4$, where $\G$ is the bubble nucleation rate. In the case in which a single field describes the transition, the bubble nucleation rate $\G$ can be computed in the semiclassical approximation using the formalism developed in \cite{Coleman:1977py, Callan:1977pt, Coleman:1980aw,Linde:1980tt,Linde:1981zj}. The confining phase transition of the WSS model involves several fields. In \cite{Bigazzi:2020phm}, as reviewed in appendix \ref{moresecWSSreview}, we took an effective approach inspired by \cite{Creminelli:2001th} where only a single field is involved. The formula for the bubble nucleation rate is reported in (\ref{Gamma2}), which involves a comparison between the efficiency of quantum and thermal fluctuations. The former are given by the $O(4)$-symmetric solution, and the latter by the $O(3)$-symmetric one.
In the analysis, we always have to verify which kind of bubble dominates.

Depending on the phase transition's efficiency, the universe may remain trapped in the false vacuum for a long time after it reaches the critical temperature, featuring supercooling. In this case, the energy density may include a temperature-independent contribution, which may start to dominate, acting as an effective cosmological constant that makes the universe inflate.\footnote{We recall, indeed, that the radiation and the vacuum contributions to the energy density scale, respectively, as $R (t) ^{-4}$ and $R (t) ^0$.} As a result, it is not guaranteed that the phase transition completes, hence in the analysis, we will always have to check that it actually does. Technically, this is done through formula (\ref{transitionok}) discussed in appendix \ref{appformule}. Depending on whether percolation enters the vacuum-dominated phase or not, the percolation temperature is computed, respectively, by (\ref{condit}) or (\ref{percolationtemperatureradiationnew}).
In performing these and the following calculations, we use the Chapman-Jouguet formula (\ref{vJC}) for the velocity of the bubble.\footnote{The friction with the plasma puts some upper bound on the velocity (see, e.g., \cite{Moore:1995si}). In our cases, an estimate of these upper bounds along the lines of \cite{vonHarling:2019gme} turns out to be always larger than the velocity calculated with (\ref{vJC}).}

Gravitational waves are produced during the propagation of nucleated bubbles in the plasma in three ways: 
collisions among bubbles,
collisions of plasma sound waves, and turbulence in the plasma.
Unfortunately, the turbulence contribution to the gravitational waves spectra is currently not well-understood. 
Typically it is deemed as subdominant. We will only consider the contributions coming from bubble collision and from the sound waves for these reasons. The formulae for the spectra in these two cases are given, respectively, by (\ref{formulascollision}) and (\ref{formulasoundwaves}).

As we discuss in appendix \ref{appformule}, it is not easy to estimate how the energy is distributed among the various contributions. Comprehension of the bubble dynamics and, most importantly, interaction with the plasma is one of the major open problems in the field so that the results are affected by huge incertitudes. For this reason, in this paper, the results for the spectra are presented separately for the bubble collision and sound waves contributions, pretending that all of the energy is concentrated in one of them in turn.  The true spectra will obviously be in between these two ``extremal" cases. 

The GW spectrum depends crucially on a parameter, usually called $\alpha$, which accounts for the amount of energy released in the transition. We are going to use its expression in terms of the trace of the energy-momentum tensor (formula  (\ref{alphacap})), adjusting in any place the number of relativistic d.o.f. at the relevant temperature scale.\footnote{Table I in \cite{Watanabe:2006qe} turns out to be a useful tool for this task.} As we will see, the spectrum with a larger magnitude is that associated with sound waves.

In the next two sections, we present the analysis in the various scenarios. From the WSS model perspective, the main difference among them is given by the choice of the parameters $M_{KK}$, $f_{\chi,L}$, $N_f$, $N$, and $\lambda$. Actually, for what concerns the  next section, the latter two enter through the combination 
\be
g \equiv \l N^2 \ .
\ee
As a general framework, although both $N$ and $\lambda$ are required to be large parameters, it is natural not to introduce a huge hierarchy of scales. Thus, we tend to prefer (but not limit ourselves) to consider not-too-large values of the parameter $g$, starting from $g \gtrsim 100$.   

\section{GWs from deconfinement/confinement phase transition}
\label{darkscenarios}
In this section, we present the GW spectra produced in three possible dark scenarios, which we name \emph{Dark HQCD 1}, \emph{Dark glueballs} and \emph{Dark axion}.
The ``H'' in HQCD stands for ``Holographic'', to underline the fact that there are extra modes w.r.t.~standard QCD-like theories. 
In these scenarios, gravitational waves are always associated with the confinement/deconfinement phase transition.
It is important to outline that the WSS model realizes explicitly, in a specific regime of parameters, scenarios which have been previously proposed in the literature (see e.g.~\cite{Bai:2013xga} and \cite{Cline:2012bz,Kribs:2016cew,Profumo:2019ujg} for reviews). While it would be very interesting to further study the phenomenological implications of this regime of parameters, in this paper we just concentrate on the gravitational waves spectra.
Thus, in the following subsections we are going to sketch the different scenarios, discussing the main information needed for the computation of the GW spectra.
The latter are determined with the formulae collected in appendix \ref{appformule} and the results are presented in subsection \ref{secresultsconfinement}.

\subsection{Dark HQCD 1}
\label{secdarkqcd}
QCD-like theories with $N_f$ flavors can provide different dark matter candidates.
Depending on the details of the models, the main fraction of dark matter can come from dark baryons, nuclei, mesons, and so on.
Analogously, the dynamically generated scale, which in the WSS model is denoted as $M_{KK}$, varies considerably among the various theories, typically from about 100 MeV to about 100 TeV.
In this subsection we consider the WSS model with $N_f$ flavors, in the regime (\ref{regimeofstudy}), as providing a strongly-correlated large $N$ dark QCD-like sector. Previous studies of gravitational wave spectra in similar scenarios include \cite{Schwaller:2015tja, Tsumura:2017knk, Aoki:2017aws, Bai:2018dxf, Helmboldt:2019pan, Aoki:2019mlt}.  
 
We have analyzed the spectra of GW produced in the phase transition for the dynamical scale values 
\be
M_{KK}=10^n \ \text{GeV} \,,\q \q \q  n=-1,0,..., 6 \ , 
\ee
and for
\be
g=10^m \ , \q \q \q m=2,3,6,10 \ .
\ee

The case $g=10^2$ is the only one where the Universe at the time of bubble percolation is in a radiation domination phase, hence we employ formula (\ref{percolationtemperatureradiationnew}) to determine the percolation temperature; in all the other cases, the Universe is in a vacuum domination era and we have to employ formula (\ref{condit}). 
For  $g=10^{2,3,6}$, the relevant bounce solution is the $O(3)$-symmetric one, while for $g=10^{10}$ the $O(4)$-symmetric configuration dominates.

In determining the reheating temperature according to formula (\ref{condtR}), care must be taken to count the correct number of degrees of freedom both in the Standard Model and in the dark sector.
In fact, in the confined phase of the dark sector there can be glueballs, KK-modes and mesons which become relativistic at the reheating temperature. This happens for $g=10^{6}$ and $g=10^{10}$.
In the first case, only the lightest glueball and KK mode must be included, together with the lightest mesons.
In contrast, in the second case, the reheating temperature is about seven times $M_{KK}$.
At this scale, many glueballs from Table 2 in \cite{Brower:2000rp} as well as many mesons must be included, giving hundreds of d.o.f.
Unfortunately, the spectrum of KK modes is not known in detail.
The first KK modes have mass of one $M_{KK}$, but we have no definite information on the number of degrees of freedom at $7 M_{KK}$.
We give a very rough estimate of this number assuming that the density of KK modes has the same dependence on the energy as the spectrum of glueballs.
We then double the number of degrees of freedom to account for the fermionic glueballs and KK modes. 
The same is done for the mesons.
However, we underline that the incertitude associated to the number of degrees of freedom introduces an error that does not spoil the order of magnitude of our results.

\subsection{Dark Glueballs}
\label{secdarkglueballs}
Another well-motivated class of dark matter candidates is represented by stable glueballs, the bound states of $SU(N)$ Yang-Mills theory. 
The WSS model with $N_f =0$ is therefore suitable for describing such a scenario and for performing in this context reliable calculations.
Being derived in the quenched approximation, the results of section \ref{secdarkqcd} can be seen as also concerning a scenario where the non-interacting dark sector is constituted by a $SU(N)$ Yang-Mills theory without flavors.

The latter can also model the case where the dark matter is actually self-interacting, a possibility which helps softening the problems of the $\Lambda$CDM model with small-scale structures \cite{Boddy:2014yra}.
In this scenario, phenomenology can be satisfied for glueball masses ranging from keV to fraction of GeV. 
When the order of the latter is around one MeV or smaller, one has to take care of phenomenological constraints related to the effective number of neutrino species and coming from Cosmic Microwave Background (CMB) measurements, and from measurements of the relative abundance of elements in Big Bang Nucleosynthesis (BBN). They imply that the dark sector cannot be in thermal equilibrium with the visible sector. In particular, the dark sector temperature $T$ has to be smaller than $T_V$, the visible sector one \cite{Breitbach:2018ddu,Fairbairn:2019xog}. As a result, non-gravitational couplings among the two sectors have to be absent or extremely small.
Whenever this is the case, gravitational waves produced in first-order transitions can be one of the few means at our disposal in order to observe direct signals coming from the dark sector. Previous studies of the GW spectra in similar cases within the context of simple effective models can be found in \cite{Breitbach:2018ddu,Fairbairn:2019xog}.  

In this section we will investigate cases with dynamical scale values
\be\label{enrange}
M_{KK}=10^n\, \text{keV} \ , \q \q n=0,1,2,3,4  \ .
\ee
The other main difference with respect to the analysis performed in section \ref{secdarkqcd} is given by the fact that the ratio $\xi = T/T_V$ can be smaller than one. We assume that $\xi$ stays constant during the bubble nucleation and GW observation process.
We explicitly explore benchmark cases where
\be
\xi=10^{-1} \ ,\q \q  g=10^{3, 5, 10} \ .
\ee
Moreover, we have considered the case where $M_{KK}=100$ keV, $g=5\cdot 10^3$ and $\xi=0.1$.
We have also checked that the smaller the value of $\xi$ is, the more the signal is suppressed.
For $\xi=10^{-5}$, for example, the signal will be completely invisible in near future facilities. 

We estimate the contraints from the CMB and the BBN on the number of relativistic degrees of freedom by parameterizing them as an extra contribution to the effective number of neutrino species $\Delta N_{eff}$ \cite{Breitbach:2018ddu}.
The constraint from the BBN, which turns out to be the most stringent one, dictates that 
$\Delta N_{eff} \lesssim 0.5$.
We use the formula  \cite{Breitbach:2018ddu}
\be\label{deltaN}
\Delta N_{eff} = \frac47 \left( \frac{11}{4} \right)^{4/3} g_*\ \xi^4 \,.
\ee
The constraint has to be imposed around $T_{V,BBN} \sim 100$ keV.
Whenever the percolation temperature is such that the dark sector is in the confining regime at $T_{V,BBN}$, $g_*$ just counts the number of relativistic glueballs and the constraint is automathically satisfied for our range of parameters because of the $\xi$ factor in (\ref{deltaN}).
If instead the dark sector is in the deconfined phase at $T_{V,BBN}$, the relevant formula for $g_*$ is, from the energy density of section \ref{seccosmo},
\be\label{gstarhere}
 g_{*} = \frac{5^2 \cdot 2^7 \pi^2}{3^6} g \frac{T^2}{M_{KK}^2}\, .
\ee
In this case large values of $g$ can overcome the $\xi^4$ suppression in (\ref{deltaN}).
In fact, for $g=10^{10}$ the constraint from the BBN is never satisfied in the range of energies (\ref{enrange}), and it restricts the allowed regimes to $M_{KK} \geq 100$ keV for $g=10^3$ (and for $g=5\cdot10^3$),  $M_{KK} \geq 1$ MeV for $g=10^5$.

Let us briefly describe the main features of the calculation of the spectra.
The decoupling of the dark and visible sectors
implies that whenever we consider plasma effects, the plasma in question is just the one of the dark sector.
As a consequence, there are two relevant $\alpha$ parameters (formula (\ref{twoalphas})), denoted as $\alpha$ and $\alpha_D$, measuring respectively the energy released in the transition w.r.t. the \emph{visible} sector energy density only and w.r.t. the \emph{dark} sector energy density only. 
The velocity of the bubble wall is determined by formula (\ref{vJC}) with $\alpha$ replaced by $\alpha_D$.
The same is true for the efficiency parameter $\kappa_v$ (formula (\ref{kappav})) for the sound wave spectra.

For $g=10^3$ (and $g=5\cdot 10^3$), in all the cases the Universe is found to be in a radiation domination era at the time of percolation.
In fact, values of $\xi < 1$ enhance the contribution of the SM energy density of radiation against the dark sector vacuum energy density.
The bubbles in these cases have $O(3)$ symmetry.
Only for the cases of $g=10^{5,10}, \xi=10^{-1}$ the Universe is in a vacuum domination era, the percolation temperature is very small due to supercooling and $O(3)$ ($O(4)$) bubbles dominate for $g=10^5$ ($g=10^{10}$).
Also, in the cases of $g=10^3, \xi=10^{-5}$ and $g=10^{10}, \xi=10^{-1}$ the reheating temperature is considerably different from $T_p$, so that we have to consider many glueball and KK modes from the dark sector.
However, due to the damping factor $\xi$, 
the contribution of the dark degrees of freedom is quite suppressed w.r.t.~the contribution of the SM particles.

\subsection{Dark Axion}
\label{secdarkaxion}

In this section, we analyze a third range of dark sector dynamical scales, relevant for composite QCD axion models.
The benchmark model is the one discussed in \cite{Kaplan:1985dv} building on the model in \cite{Kim:1979if} (see also \cite{Choi:1985cb}, and \cite{DiLuzio:2020wdo,Lombardo:2020bvn} for recent reviews).

In its simplest realization, the model comprises a dark $SU(N)$ Yang-Mills sector and four massless flavors in its fundamental representation.  Three of them form a triplet of the QCD $SU(3)_c$ gauge group, whereas the fourth constitutes a singlet. The global symmetry includes an axial $U(1)_A$, which plays the role of the Peccei-Quinn symmetry. In fact, the latter is anomalous and spontaneously broken by the flavor condensation due to the strong dynamics of the dark $SU(N)$. The associated pseudo-Nambu-Goldstone boson is then a composite axion.
In this scenario, the confinement/deconfinement transition of the dark $SU(N)$ theory implies the Peccei-Quinn phase transition, which is of the first order. Previous studies of GW spectra from Peccei-Quinn transitions in effective theories (possibly of bottom-up Randall-Sundrum type) can be found in \cite{Croon:2019iuh,vonHarling:2019gme,DelleRose:2019pgi}.

In the model of  \cite{Kim:1979if}, the axion decay constant $f_a$ is related to 
$f_\chi$ by
\be
f_a = \frac{\sqrt{6}}{N} f_\chi \ .
\ee
Thus, from (\ref{decayconstantpion}), we read
\be
f_a = \frac{1}{3 \pi^2} \sqrt{\frac{\lambda}{2 N}}M_{KK}\ . 
\ee
Consistency with phenomenology requires $f_a \gtrsim 10^8$ GeV. Moreover, formula (\ref{epsilonf}) with $N_f=4$ gives the constraint
\be \label{lambdapic}
\lambda \lesssim 3 \sqrt{N}\ ,
\ee  
and therefore we are led to consider dynamical scales  $M_{KK} \gtrsim 10^9$ GeV.
We will consider two benchmark values of $g$,
\be
g= 10^3 \ , \q \q \q g = 10^8 \ .
\ee

The details of the calculations are very similar to the ones in section \ref{secdarkqcd}.
In all the cases, the Universe is in an energy domination era at the time of percolation.
For $g=10^3$ ($g=10^8$) the $O(3)$ ($O(4)$) bounce dominates.
In order to determine the reheating temperature for $g=10^8$, we have to take into account glueball, KK and mesonic degrees of freedom.

\subsection{Results for the spectra}
\label{secresultsconfinement}

In this section, we describe the results for the GW spectra generated by the first-order confinement/deconfinement transition of the holographic model. As we have already mentioned, we do not consider the contribution from turbulence in the plasma and we separately consider the contributions from bubble collisions and sound waves.

For what concerns the sound waves contribution, there is a further incertitude due to the unknown source duration.
Until very recently, the source was expected to last for a long time in Hubble units. 
Under this assumption, most of the literature has employed the formulae reviewed in \cite{Caprini:2015zlo,Weir:2017wfa}.
However, it has been recently pointed out that the source can be quite short, see e.g. \cite{Ellis:2018mja, Ellis:2019oqb, Ellis:2020awk,Caprini:2019egz,Guo:2020grp}. Accordingly, the power spectrum is quenched by the short time factor (\ref{factor}). 
In this paper, an agnostic attitude is taken and both spectra, with and without quenching factor, are presented.
This allows us to have an idea of the possible range of the signal and to compare the results with previous literature.

In summary, three types of spectra are calculated: the one from bubble collisions $\Omega_c$, the one from sound waves without quenching factor $\Omega_{sw}$ and the one from sound waves with quenching factor $\Omega_{sw,q}$.
As a general trend, $\Omega_c$ is found to give the smallest peak signal.
Moreover, the peak frequency increases with $M_{KK}$ and the amplitude of the signal increases with $g$.

\begin{figure} 
\center
\includegraphics[scale=1.6]{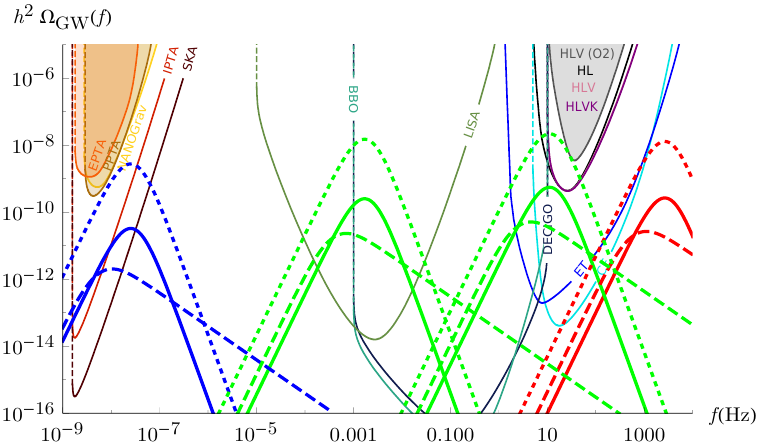}
\caption{Examples of GW power spectra $h^2\Omega_{GW}$ due to bubble collisions ($\Omega_c$, dashed lines) and sound waves in the case of \emph{short} source duration ($\Omega_{sw,q}$, continuous lines) and long source duration ($\Omega_{sw}$, dotted lines). Expected sensitivities (PLISCs) for a number of experimental facilities are reported for comparison \cite{Schmitz:2020syl}. From left to right, the spectra correspond to the following parameters: $(M_{KK}/{\rm GeV}, g)=(10^{-4},5\cdot 10^{3})$ (blue lines), $(10^2,10^6), (10^6,10^6)$ (green lines), $(10^9,10^3)$ (red lines).}
\label{figtutti}
\end{figure}

In figure \ref{figtutti} we report examples of power spectra.
In the plot, a few benchmark values of the parameters $M_{KK}, g$ are chosen to show the detectability potential of the GW emissions.
A number of experimental sensitivities are shown for comparison.

The first clear result is that in various cases the GW signals are going to be detectable in near future experiments, with the possible exception of the composite axion model.

Notice that $\Omega_c$ and $\Omega_{sw,q}$ approximately span an order of magnitude in power of the signal around the peak, represented in the figure by the regions in between the dashed and continuous curves. Notice, moreover, that the upper value of the signal for the $\Omega_{sw}$ spectra (dotted lines) is greatly amplified w.r.t. the quenched case $\Omega_{sw,q}$ (continuous lines); the true signal from sound waves is expected to be in between the two types of lines.
The total signal is expected to be a combination of the one from sound waves and the one from collisions.

The blue lines at the left of the plot show a representative case for a small dynamical scale value, $M_{KK}= 100$ keV, relevant for the Dark Glueballs scenario, for $g=5 \cdot 10^{3}$ and for the value $\xi=0.1$ of the ratio between the dark and the visible sector temperatures.  
It is clear that the signal is potentially detectable by pulsar timing array experiments such as IPTA and SKA. Actually, the most ``optimistic'' scenario where almost all the energy of the process goes into GWs from sound waves is of great experimental interest. In fact, in this case the signal could be visible in current single experiments such as NANOGRAV, EPTA and PTTA.  Actually, very recently, the results of 12.5 years observations by NANOGRAV have been reported in \cite{nanograv}, showing strong evidence for a stochastic spectrum compatible with GW signals with frequency peak around $10^{-9}-10^{-8}$ Hz and average energy density $\langle h^2\Omega_{GW}\rangle\sim 10^{-10}$. If, among the possible sources of this signal, there is space for a cosmological strongly first-order phase transition in a dark sector - as it has been recently suggested in \cite{Nakai:2020oit, Addazi:2020zcj, Ratzinger:2020koh} - our Dark Glueball model could be viewed as a possible candidate.

Although it is not shown in the figure, the same possibility of detection happens if $g=10^{3,5}$ (again for $\xi=0.1$) for $M_{KK}$ around $0.1 - 1$ MeV, at least in the SKA experiment.

The two sets of green lines at the center of the plot correspond to the parameter value $g=10^6$ and energies respectively of $M_{KK}=10^{2}$ and  $M_{KK}=10^{6}$ GeV, relevant for the Dark HQCD 1 scenario. The first case is going to be detectable already by LISA and clearly by the more sensitive experiments such as BBO and DECIGO. The same remains true down to $M_{KK} \sim 10$ GeV and $g \sim 10^2$ (not shown in the plot).
The second case of $M_{KK}=10^{6}$ GeV is detectable by ET or CE facilities.
Of course, all the intermediate energies can be detected, and this remains true even for smaller values of $g$ down to $10^2$ and larger values of $M_{KK} \lesssim 10^7$ GeV.
For $g = 10^{10}$ the signal is visible at LISA starting from $M_{KK} \sim 1$ GeV.
Thus, a few near future experiments (LISA and ET for example) are going to be able to fully probe strongly coupled dark QCD-like sectors (with large ranks) in the energy range $M_{KK} \sim 1-10^7$ GeV.

Finally, the three red lines at the right of the plot correspond to $g=10^3$ and $M_{KK}=10^9$ GeV, and are relevant for the Dark Axion scenario with $f_a \sim 10^8$ GeV. Only in the optimistic case in which the duration of the sound waves' source is long, the spectrum falls within the sensitivity curve of CE. Since we expect the real signal to be in the region between the three curves, this case is unlikely to be detectable in near-future experiments. Moreover, if $M_{KK}$ increases, such that $f_a > 10^8$ GeV, the curves are shifted to larger values of the peak frequencies. As a result, the Dark Axion scenario is not favorable for producing detectable gravitational waves.

\begin{figure} 
\center
\includegraphics[scale=0.99]{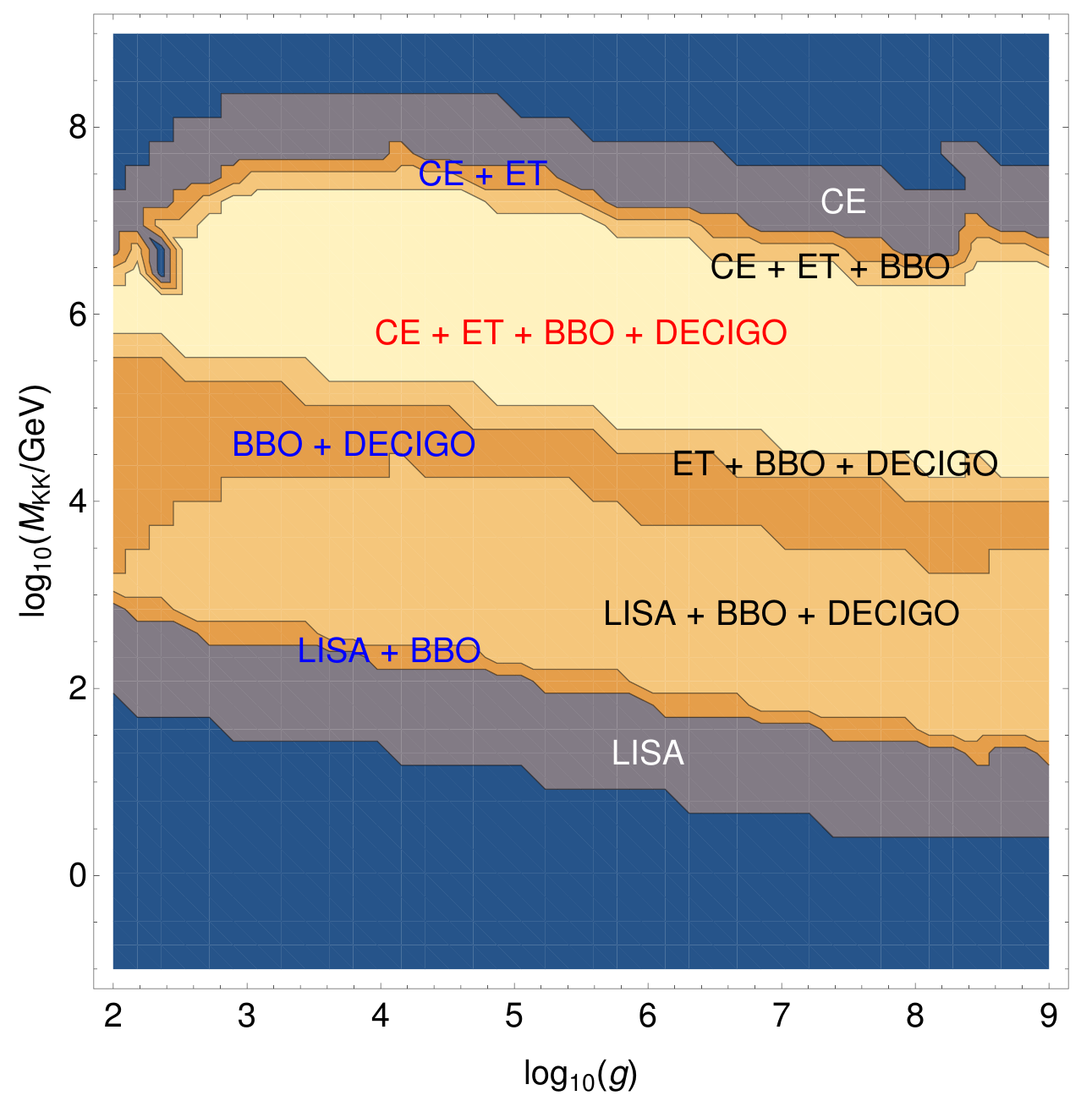}
\caption{Rough estimate of the possibility of detection in future facilities of GWs produced by holographic
 first-order confinement phase transitions.
The plot explores the parameter space of the dark HQCD 1 and dark axion scenarios and considers GWs
 produced by sound waves, eq.~(\ref{formulasoundwaves}), including the suppression factor (\ref{factor}). 
 The color code indicates the
number of facilities that could measure the signal for a particular value of the parameters: none (blue),
one (grey), two (dark orange), three (light orange) or four (yellow).}
\label{newfigure}
\end{figure}

Figure \ref{newfigure} illustrates some of the results, depicting the regions of parameter space that could be explored by five 
facilities projected for the near future (CE, ET, BBO, DECIGO and LISA). 
The current capabilities of LIGO and VIRGO are insufficient
for detection, although they come quite close for $10^6$ GeV $< M_{KK}< 10^7$ GeV and $g>10^4$ and, therefore, these facilities and KAGRA have been
left out of the plot. In the figure, only the dark HQCD 1 and the dark axion scenarios are considered because the dark glueballs model would require
introducing the extra parameter $\xi$. As a benchmark case, we have chosen to make the plot using the predicted spectrum of GWs produced by sound waves, taking
into account the suppression factor due to short pulse duration. For the plot, we just consider the spectrum at the frequency $f_{det}$ at which each detector attains
its best sensitivity and compare it to $h^2 \Omega_{sw}(f_{det})$. This is certainly a simplification which, together with all the approximations 
and assumptions involved in the derivation of $h^2 \Omega_{sw}(f)$, implies that the contours of the figure should be considered only as very rough estimations.

However, the picture that emerges is clear: facilities in the near future should be able to investigate 
the GW spectrum stemming from large regions of the parameter space of holographic theories with a first-order phase transition in the early Universe.
Moreover, in various (optimistic) scenarios, stochastic GW background generated in this type of models can be detectable by the advanced version of currently running experiments.  
Large $M_{KK}$ is probed by devices that concentrate in large GW frequencies. In fact, the small values of $M_{KK}$ of the dark glueballs scenario can
only be measured with detectors of small frequency GWs such as pulsar timing arrays, as shown in figure \ref{figtutti}. The dependence on the coupling of the gauge theory $g$ is only mild, provided that  it is large enough for the holographic description to apply.

\section{GWs from chiral phase transition}
\label{sec:chisb}

In this section, we consider scenarios that display a chiral symmetry breaking/restoration phase transition separated from the eventual confinement/deconfinement one. This implies the fascinating consequence of having two distinct peaks in the spectrum of stochastic GWs.
Firstly, we discuss the possible scenarios and then we present the results for the spectra.

\subsection{Dark HQCD 2}
\label{secdarkqcd2}
The scenario that we consider in this subsection is a close cousin of the Dark HQCD 1 scenario of subsection \ref{secdarkqcd}: the WSS model describes a dark sector, very weakly interacting with the Standard Model (in the most extreme case, even interacting with the Standard Model only gravitationally).
The difference with respect to what has been discussed in section \ref{secdarkqcd} concerns the choice of the WSS parameter $L$. In section \ref{secdarkqcd}, the latter was taken to be $L=\pi M_{KK}^{-1}$, corresponding to the chiral symmetry breaking scale $f_{\chi}$ given in (\ref{decayconstantpion}). In contrast, here we will consider cases with $L\ll\pi M_{KK}^{-1}$, for which the chiral symmetry breaking scale $f_{\chi,L}$ is given in (\ref{fchil}). 
As said, this implies that the chiral phase transition is separated from the confinement one.

An important difference with respect to the scenario of section \ref{secdarkqcd} is that the evolution of the Universe cannot be considered to be adiabatic from the time of the chiral symmetry breaking transition to the present time, since there is a second first-order phase transition. 
This calls for a correction of the standard formulae for the redshift of the signal,
which are derived under the assumption of adiabatic evolution. 
In fact, the adiabatic assumption holds from the time of the chiral symmetry breaking transition
to the percolation time of the confinement transition.
Then, assuming fast reheating in the confinement transition, the temperature has a sudden jump from the percolation temperature to the reheating temperature.
Finally, from this time to the present day, the Universe continues to evolve adiabatically. 
In appendix \ref{sec:redshift} this behavior is reflected in formulae (\ref{fd2}), (\ref{Hd2}) for the frequency and power spectrum redshifts. 

A consequence of these formulae is that the magnitude of the chiral symmetry breaking transition signal decreases if the value of $g= \lambda N^2$ increases. 
This is due to powers of the ratio of the percolation and reheating temperatures of the confining transition, $T_{p,conf}, T_{R,conf}$, appearing in formulae (\ref{fd2}), (\ref{Hd2}) (in a coefficient which we have called $\delta$ in (\ref{delta})).
As we semi-analytically estimate in appendix \ref{estimates}, an increase of $g$ implies more supercooling, hence $T_{p,conf}$ decreases and at the same time $T_{R,conf}$ increases, resulting in a suppression of the GW signal. 
For this reason, in the present scenario we are going to describe the case where $\l$ and $N$ are such that $g$ has a ``\emph{small}'' value. 
In particular, we will investigate the representative case
\be
\label{choicedarkQCD2}
\lambda=N=10 \ , \q \q \q g = 10^3 \ .
\ee

It is convenient to introduce dimensionless quantities,
\be
\tilde f_{\chi} \equiv \frac{f_{\chi,L}}{M_{KK}} \ , \q \qquad \tilde T \equiv \frac{T L}{0.1538} \sim 0.35 (\lambda N)^{1/3} \frac{T}{M_{KK}^{1/3} f_{\chi,L}^{2/3}}\ ,
\ee
such that the critical temperature for the chiral symmetry breaking transition corresponds to $\tilde T =1$.
The condition that the chiral symmetry breaking transition happens above the deconfinement transition gives the constraint
\be
\tilde f_{\chi} > 0.013 \lambda^{1/2} N^{1/2}\ ,
\ee
that with the choice (\ref{choicedarkQCD2}) corresponds to $\tilde f_{\chi} > 0.13$. 

In fact, the signal is enhanced if the chiral symmetry breaking scale $\tilde f_{\chi}$ is large. The validity of the quenched approximation we are assuming for the flavors constrains the magnitude of this parameter. In particular, the requirement that the approximation works at the percolation temperature and at the reheating temperature sets the limit $\tilde f_{\chi} \leq 60$ for the choice of parameters (\ref{choicedarkQCD2}).
This comes from the requirement that the energy density of the flavors is subleading with respect to the one of the gluonic degrees of freedom, see section \ref{seccosmo}. 
Thus, we will consider the benchmark values
\be
\tilde f_{\chi} = 30  \ , \q \q \q \tilde f_{\chi} = 60 \ .
\ee

A noticeable difference with respect to the cases analyzed in section \ref{darkscenarios} is that the energy released in the transition is much smaller than the energy of radiation, since the former comes from the flavors, which are quenched, while the latter mostly comes from the gluons. 
As a result, the parameter $\alpha$ is much smaller than one in this case and
 the bubble velocity sometimes is not very close to unity.
Since the energy released in the transition is small as compared to the total energy, we expect the reheating temperature to be close to the percolation temperature.

Regarding the counting of degrees of freedom,
in the case at hand, by normalizing the entropy density as 
\be
s = \frac{2 \pi^2}{45} g_{*}^S T^3 \ ,
\ee 
at the time of emission we have the three contributions from the Standard Model, gluons and flavors
\be
g_{*}^S = g_{*,SM} + g_{*,glue}^S + g_{*,\chi}^S\ ,
\ee
with \cite{Aharony:2006da}
\begin{subequations}
\bea
 g_{*,glue}^S &=& \frac{5 \cdot 2^6 \pi^2}{3^4}\lambda N^2 \frac{T^2}{M_{KK}^2}\ , \\
 g_{*,\chi}^S &=& \frac{5 \cdot 2^4}{3^6}\lambda^3 N_f N \frac{T^3}{M_{KK}^3}\ .
\eea
\end{subequations}
From the energy density of section \ref{seccosmo} we read
\be
g_{*} = g_{*,SM} + g_{*,glue} + g_{*,\chi}\ ,
\ee
with
\begin{subequations}
\bea
 g_{*,glue} &=& \frac{5^2 \cdot 2^7 \pi^2}{3^6}\lambda N^2 \frac{T^2}{M_{KK}^2}\ , \\
 g_{*,\chi} &=& \frac{5 \cdot 2^7}{7\cdot 3^6}\lambda^3 N_f N \frac{T^3}{M_{KK}^3}\ .
\eea
\end{subequations}

\subsection{Holographic Axion}
\label{secholoaxion1}
Another scenario where a chiral symmetry breaking/restoration takes place is the holographic QCD axion model of \cite{Bigazzi:2019eks}, which we call \emph{HoloAxion} in the following. 
The WSS theory is considered as a model for the strong interactions of the Standard Model, including the QCD axion physics. 
The axion arises as a composite particle, analogous to the $\eta'$, coming from an extra flavor with $L \ll \pi M_{KK} ^{-1}$ so that it condenses at a large scale $f_a=f_{\chi,L} \gg \Lambda_{QCD}$. In contrast, the SM quarks are embedded in such a way that the related chiral symmetry breaking scale is given by $f_{\chi}$ in (\ref{decayconstantpion}). The condensation of the axion is a Peccei-Quinn first-order transition which can therefore generate gravitational waves.

The energy density of the false vacuum configuration in this case reads formally as (\ref{energytot}).
Let us briefly comment on each contribution. Since the QCD sector of the theory, gluons and quarks, is described by the WSS model, the related relativistic degrees of freedom are not counted in $g_{*}^{SM}$ (which then has $27.75$ as its maximal value) in $\r_{rad, SM}$. Concerning $\r_{rad, \chi}$, the number of flavors in (\ref{energyflavorconfinedantipodal}) is $N_f=7$, because we have six QCD quarks plus an extra flavor that provides the axion. The contribution $\r _{0, \chi}$ is given by (\ref{energyflavorconfinedantipodal2}) with $N_f =6$, because the latter holds only for the case $L = \pi M_{KK} ^{-1}$. The remaining flavor gives a contribution analogous to (\ref{energyflavorconfinedantipodal2}) but suppressed by a factor of $M_{KK}/f_{\chi,L}$, hence it can be neglected.

Since in this scenario the WSS model describes the strong sector, the usual, uncontrolled extrapolation of the regime of validity of these formulae to the real world parameter values is performed.
This amounts to quitting the planar regime by setting $N=3$.
Then, the parameters $\lambda$ and $M_{KK}$ are determined by fitting the $\rho$-meson mass and the value of $f_{\pi}=f_{\chi}$, giving \cite{Sakai:2004cn}
\be
\lambda = 33.26\ , \qquad M_{KK} = 0.949\ {\rm GeV}\ . 
\ee
The probe approximation is also dropped in this regime of parameters, as usual in the WSS model.
The choice of the parameter $L$ which sets $f_a$ is constrained by the requirement
\be
10^8 \ \text{GeV} \lesssim  f_a \lesssim 10^{17} \ \text{GeV} \ .
\ee
coming from axion phenomenological constraints.

\subsection{Results for the spectra}
\label{secresultschiral}
Let us comment on the behavior of the spectra that we find in the scenarios where a chiral symmetry breaking/restoration transition occurs.

In the Dark HQCD 2 scenario, two separated phase transitions occur and the signal is given by the sum of the signals of the two phase transitions. Since we work in the quenched approximation (\ref{epsilonf}), the chiral symmetry phase transition
is characterized by smaller released energies and therefore smaller signal magnitudes with respect to the confinement/deconfinement one.
The peak of the signal of the chiral symmetry transition is at higher frequencies than that due to the confinement/deconfinement transition. Being smaller, the former might be negligible with respect to the tail of the confinement signal and therefore the chiral symmetry phase transition would be effectively unobservable.\footnote{Indeed, we recall that the formulae for the spectra are affected by the incertitudes mentioned in the introduction, hence a big chiral symmetry signal is needed in order to be significant.} Since the signals associated to bubble collisions are suppressed with respect to the ones due to sound modes, we discuss only the latter.

\begin{figure} 
\center
\includegraphics[scale=0.42]{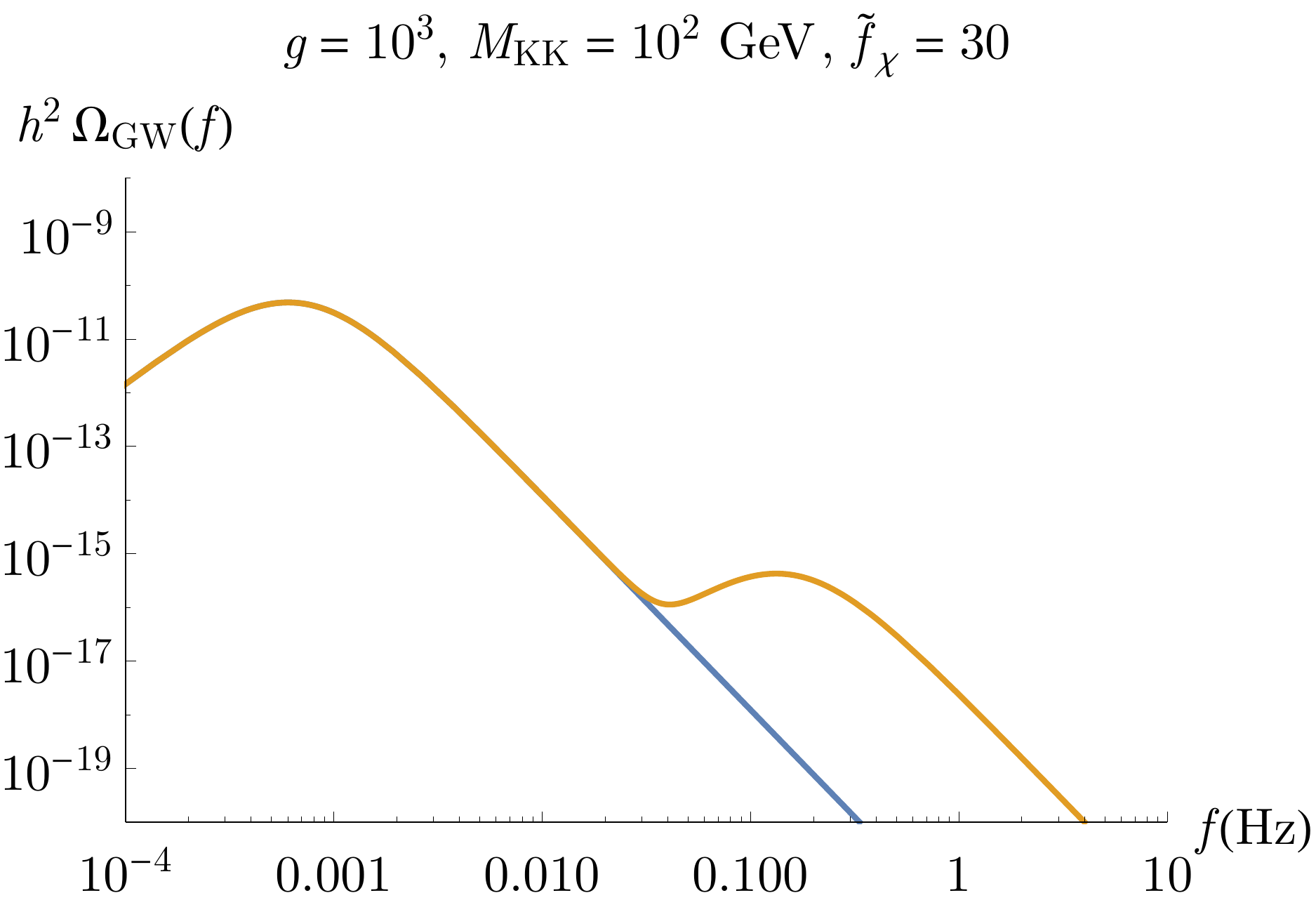}
\includegraphics[scale=0.42]{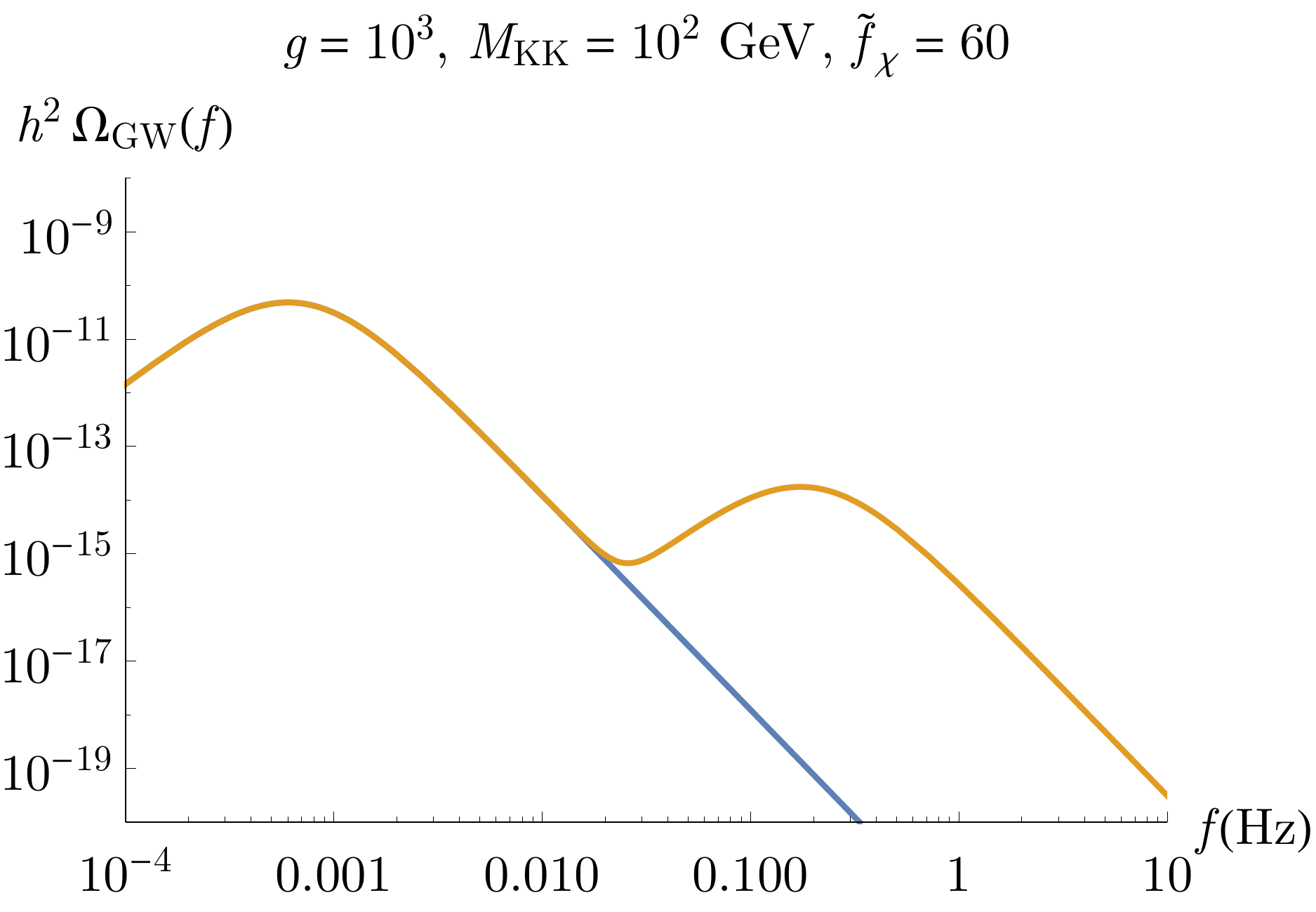}

\vspace{2mm}
\includegraphics[scale=0.42]{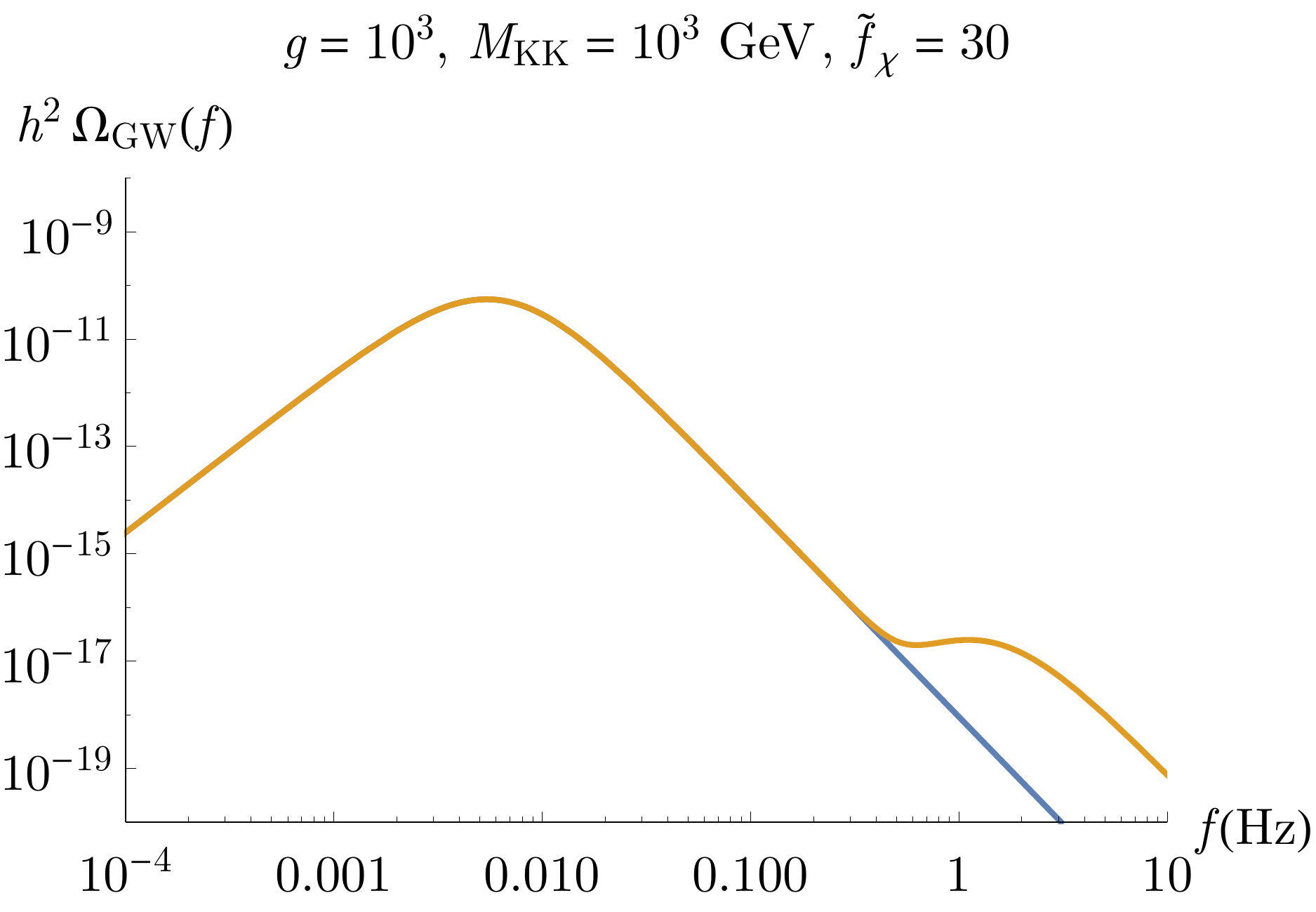}
\includegraphics[scale=0.42]{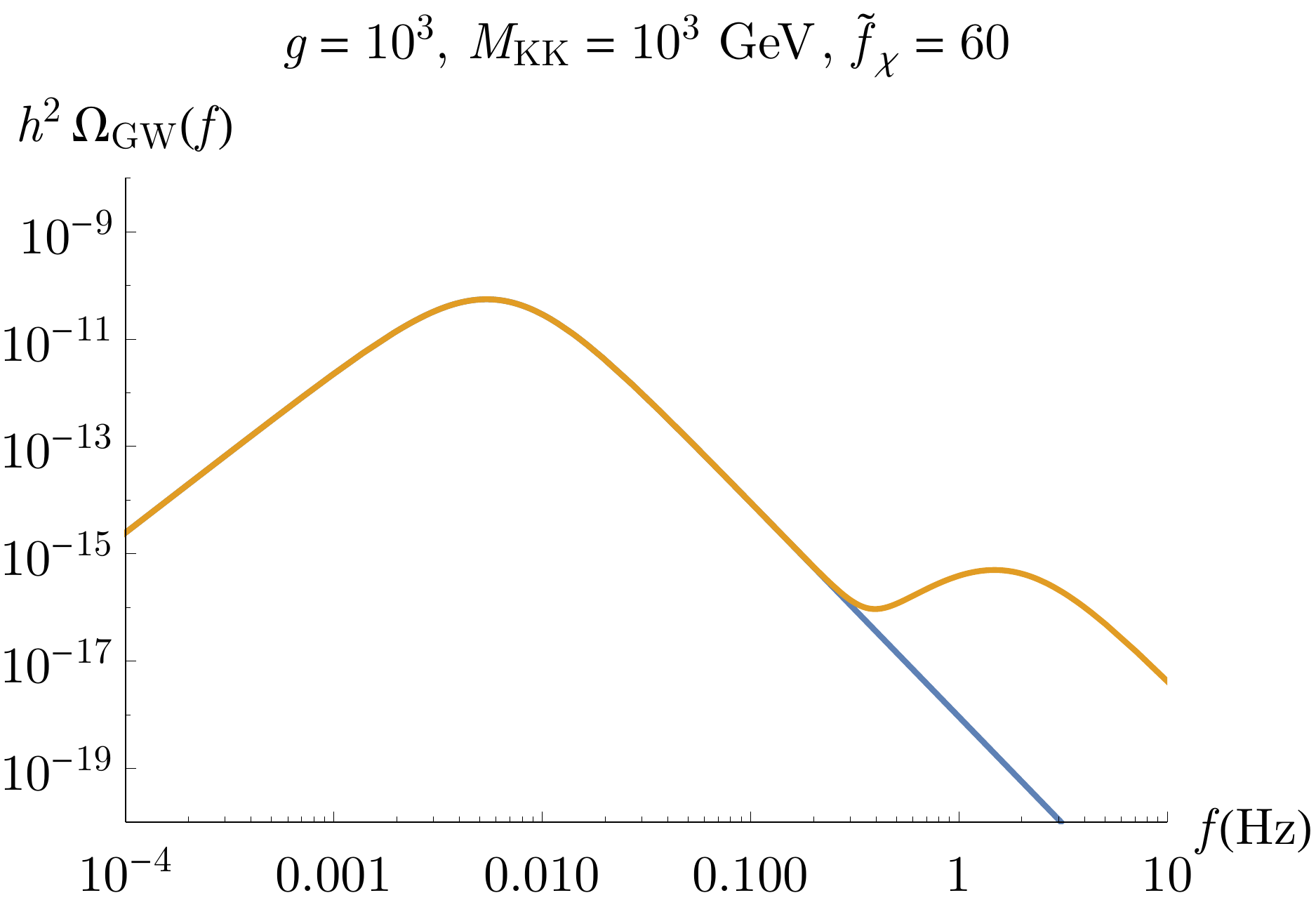}
\caption{The GW power spectra from the confinement transition (blue lines, sound modes) and from the sum of the confinement and chiral symmetry transitions (orange lines, sound modes), for different values of the parameters.
The spectra on the first (second) line are calculated without (with) the correction factor (\ref{factor}) for the chiral transition.
}
\label{figYES}
\end{figure}

Examples of the signals for different values of the parameters, with and without the correcting factor (\ref{factor}) for the duration of the transition, are reported in figure \ref{figYES}. 
Clearly, larger values of $\tilde f_{\chi}$ are more effective in separating the peak due to the chiral symmetry transition from that due to confinement.

\begin{figure} 
\center
\includegraphics[scale=1.4]{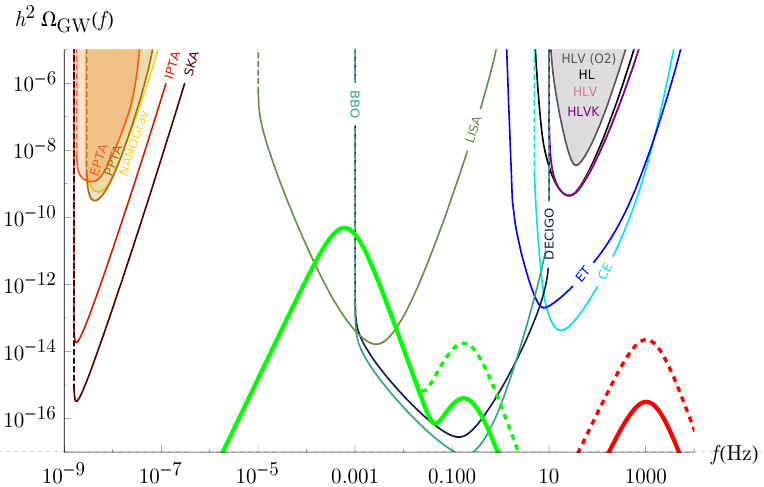}
\caption{Experimental sensitivity curves (PLISCs from \cite{Schmitz:2020syl}) and examples of theoretical GW power spectra from sound waves. In green, the sum of the signal from the confinement transition with $\lambda=N=10, M_{KK}=100$ GeV and that from the chiral symmetry transition with $N_f=1, \tilde f_{\chi}=60$. In red, the spectra for the HoloAxion case with $f_a \sim 10^8$ GeV. Continuous (dashed) curves correspond to the signal with (without) suppression factor (\ref{factor}) for the short duration of the chiral transition.}
\label{experimentchirallittle}
\end{figure}

Figure \ref{experimentchirallittle} offers an example of the scenarios that we have been discussing in this section, presenting the comparison of the computed spectra with the sensitivity curves of experiments. The green curves correspond to a representative two-peak case in the Dark HQCD 2 scenario, namely that where $M_{KK} = 100$ GeV and $\tilde f_\chi = 60$. 
It displays a large peak due to the confinement/deconfinement transition at frequency $f \sim 10^{-3}$ Hz which fits into the sensitivity curve of LISA, and a smaller peak due to the chiral symmetry transition at frequency $f \sim 10^{-1}$ Hz which does not fit into the LISA sensitivity curve but is expected to be visible by the next generation facilities such as BBO and DECIGO. The conclusion is that the two-peak signal is certainly within reach of the next generation facilities at least for a certain region of parameter space. 

Concerning the HoloAxion case, the result for the extremal case where the axion decay constant takes the lower allowed value $f_a \sim 10^8$ GeV is displayed in red in figure \ref{experimentchirallittle}.
The frequencies of the peak of these curves are too large and their magnitudes are too small to be captured by near-future facilities like ET or CE, even in the optimistic case in which we do not include the quenching factor  (\ref{factor}) due the duration of the sound waves.
Moreover, as $f_a$ increases, the peak frequencies increase as well, hence going further away from the sensitivity curves of the experiments. 
Thus, we conclude that the Peccei-Quinn transition in the HoloAxion scenario cannot be seen in near future experiments.

\section{Conclusions}
\label{secconclusions}

Cosmological first-order phase transitions generate stochastic gravitational wave backgrounds potentially visible in present and next generation experimental facilities.
Dark sectors, as hidden sectors interacting with the standard model very weakly, are good candidates where to explore such transitions.

Many dark sector models in the literature are Yang-Mills or QCD-like theories.
If the rank of the gauge groups of these theories is sufficiently large, the planar limit constitutes a good approximation to their dynamics.
In this paper, we have considered the scenario where a dark sector admits a top-down holographic dual description in the gravity regime.
This means that the theory is in the planar limit and there is a gap in the spectrum of hadron masses.
When the theory admits such a dual description, we have full control on its strongly-coupled dynamics, without the need to employ effective models and uncontrolled approximations.\footnote{See \cite{Croon:2020cgk} for a discussion of the uncertainties associated to the perturbative approach.}
But even if the theory is not exactly in this regime, one can view the holographic description as an effective tool to model the strong coupling dynamics - this latter approach has been used extensively for QCD.

Describing dark sectors by means of dual gravitational theories opens up the possibility of studying their dynamics at strong coupling.
In this paper, we concentrated on the production of gravitational waves in first-order transitions.
Using the well-known Witten-Sakai-Sugimoto holographic model, we have investigated two types of transitions.
The first type is the confinement transition, possibly implying a chiral symmetry breaking transition.
The second type is a chiral symmetry breaking transition separated from the confinement one - the latter happening at a later time in the cosmological evolution.

Making use of the bubble configurations studied in the companion paper \cite{Bigazzi:2020phm}, we have been able to calculate all the relevant parameters necessary for the determination of the GW spectra.
The latter are usually affected by a number of assumptions and sometimes uncontrolled approximations.
The holographic approach allowed us to erase from this number the use of uncontrolled approximations to the strong dynamics of the dark theory.

The results of our investigation are partially in line with other studies in the literature.
In table \ref{table} we report the benchmark cases displayed in figures  \ref{figtutti} and \ref{experimentchirallittle}. 
In the case of the single confinement transition, there is a large part of the parameter space of the theory where the GW signal is going to be detectable in the next generation facilities (see figures \ref{figtutti} and \ref{newfigure} for examples).
These include pulsar timing arrays as well as space- and ground-based interferometers, depending on the dynamical scale of the theory.
Interestingly, a window of parameter space can produce a signal within the current NANOGrav sensitivity, explaining the recent potential observation in this experiment.

\begin{table}[h!]
\begin{center}
\begin{tabular}{|l|c|c|c|}
 \hline
 \multicolumn{4}{|c|}{Summary of the benchmark cases} \\
 \hline
Scenario & Dynamical scale & Chiral scale & Experiment\\
 \hline
Dark HQCD 1 & $10^2, 10^6$ & $10^2, 10^6$ & LISA-BBO, ET-LIGO \\
Dark Glueballs & $10^{-4}$  & - & NANOGrav-IPTA-SKA\\
Dark Axion  & $10^9$  & $10^8$ & (ET-CE)\\
Dark HQCD 2 & $10^2$  & $6\cdot 10^3$  & LISA-BBO-DECIGO \\
HoloAxion & 0.949  &  $10^8$  & - \\
\hline
\end{tabular}
\end{center}
\caption{Values of the dynamically generated scale and the chiral symmetry breaking scale (or axion decay constant) of the WSS model for the benchmark cases considered in figures \ref{figtutti} and \ref{experimentchirallittle}. In the last column we report some experiments with the potential of detecting the corresponding signals. In the Dark Axion case the experiments are in brackets because the detectability is marginal.
All the energies are expressed in GeV.}
\label{table}
\end{table}

When the chiral symmetry breaking transition is separated from the confinement one, the model predicts two distinct peaks in the GW spectra.
Detection of both peaks would represent an exciting smoking gun for the models with two transitions. 
The gravity regime allows to explore faithfully a branch of parameter space where the chiral symmetry signal is smaller than the confinement one.
Nevertheless, we have shown that there are certain values of parameters allowing for observation of the two peaks, for example by space-based interferometers (figure \ref{experimentchirallittle}). 
It would be interesting to study the correlations of the two peaks, which could distinguish the holographic model from other models with two phase transitions.

Finally, we have considered Peccei-Quinn transitions in two distinct axion models: a standard composite axion from a hidden sector \cite{Kim:1979if,Kaplan:1985dv,Choi:1985cb} and the recently introduced holographic axion model \cite{Bigazzi:2019eks}.
Unfortunately, in both cases, the lower bound on the axion decay constant around $10^{8}$ GeV corresponds to a peak frequency which is too large for detection in the near future. 
In this respect, the model is distinct from the holographic bottom-up (phenomenological) ones recently investigated in \cite{DelleRose:2019pgi,vonHarling:2019gme}, where the possibility of tuning a very small parameter, measuring the departure from conformality, allows to produce signals within the sensitivity of ET or CE.

In this paper we have started to
use top-down holographic models to study dark (hidden) sectors.
It will be clearly interesting to employ this approach to first characterize the model parameter space compatible with current observational constraints, and then produce predictions for observables in the strong coupling regime of the theory.


\vskip 15pt \centerline{\bf Acknowledgments} \vskip 10pt 

\noindent 
We are indebted to Riccardo Argurio, Chiara Caprini, Paolo Creminelli, Luigi Delle Rose, Alberto Mariotti, Alberto Nicolis, Diego Redigolo and Andrea Tesi for invaluable insights. We also thank Catia Grimani, Gianluca Maria Guidi, Elena Vannuccini for useful references on various experimental setups. A.L.C. would like to thank the Galileo Galilei Institute for theoretical physics and the ULB for their hospitality during the preparation of this work. A.C. thanks the ULB for the kind hospitality during the preparation of this work. The work of A.P. is supported by grants ED431B 2018/57 from Conselleria de Educacion, Universidade e Formacion Profesional and FIS2017-83762-P from Ministerio de Economia, Industria y Competitividad, Spain.


\appendix
\section{More on the Witten-Sakai-Sugimoto model}
\label{moresecWSSreview}
In this appendix, we review the holographic construction of the Witten-Sakai-Sugimoto model employed in the present paper.

\label{secWSSreviewfeatures}

Let us briefly describe the string theory embedding of the WSS model. A stack of $N$ D4-branes wrapped on a circle $S_{x_4}^1$ with coordinate $x_4 \sim x_4 + 2 \pi / M_{KK}$ give rise to the fields that transform in the adjoint representation of the gauge group. Fields transforming in the fundamental representation of the gauge group are introduced through pairs of D8/anti-D8-branes. These are transverse to the circle $S^1_{x_4}$ in such a way that the D8-branes and anti-D8-branes are separated by a distance $L \leq \pi M_{KK}^{-1}$ along $S^1_{x_4}$. 
When the $N_f$ fundamental fields are massless, the theory exhibits a $U(N_f)_L\times U(N_f)_R$ classical global symmetry, the chiral symmetry, which is realized by the gauge symmetry of the D8/anti-D8-branes. When $L = \pi M_{KK}^{-1}$, the scale of chiral symmetry breaking coincides with the confinement scale. This is the choice of parameters that is useful to model QCD, and that was, indeed, considered in the original version of the model \cite{Sakai:2004cn}. In the general case, $L$ can be considered as a free parameter of the model. This latter case has been considered in the recently proposed Holographic QCD axion scenario \cite{Bigazzi:2019eks,Bigazzi:2019hav}.

In the regime (\ref{regimeofstudy}), the gauge plus matter adjoint sector can be studied by considering the near-horizon limit of the backreaction of the D4-branes, that is a solution of Type IIA supergravity dual to a theory known as Witten-Yang-Mills (WYM). It is given by a curved metric, a non-trivial dilaton, and a four-form Ramond-Ramond (RR) field strength. If we consider the theory at finite temperature, the time direction is compactified on the circle $t \sim t + 1/T$, and therefore we have two circles: $S^1 _t$ and $S^1 _{x_4}$. As a result, depending on the temperature, there are two competing solutions. Let us briefly present them, working with the Euclidean signature.

The background that dominates in the high-temperatures regime is the black hole one,
\bea
\label{wittenbackground}
ds^2 &=& \pr{\frac{u}{R}}^{3/2} \pq{f_T(u) dt^2 + dx^i dx^i +  d x_4 ^2 } +\pr{\frac{R}{u}}^{3/2} \pq{\frac{du^2}{f_T(u)} +u^2 
d \O _4 ^2}\ ,\nonumber \\
f_T(u) &=&1- \frac{u_{T} ^3}{u^3}\ , \quad e^\ff=g_s \pr{\frac{u}{R}}^{3/4}\ ,\quad F_4= \frac{3 R^3 }{g_s} \o_4\ , \quad R^3=\pi g_s N l_s ^3\ ,
\eea
where $\omega_4$ is the four-sphere volume form and $g_s, l_s$ are the string coupling and string length.
The parameter $u_T$ is related to the Hawking temperature $T$ by
\be
u_T = \frac{16 \pi ^2}{9} R^3 T ^2 \ .
\label{utth}
\ee
The background that dominates in the low-temperature regime is called ``solitonic" and reads
\bea
\label{solitonicW}
ds^2 &=& \pr{\frac{u}{R}}^{3/2} \pq{dt^2 + dx^i dx^i + f(u) d x_4 ^2 } +\pr{\frac{R}{u}}^{3/2} \pq{\frac{du^2}{f(u)} +u^2 
d \O _4 ^2}\ ,\nonumber \\
f(u) &=& 1 -\frac{u_0 ^3}{u^3}  \ , \q \q \q u_{0} = \frac{4}{9} R^3 M_{KK}^2 \ ,
\eea
with the dilaton and $F_4$ fields keeping precisely the same form as in the previous case. 
The holographic dictionary that relates the string theory quantities and the field theory ones reads
\be
g_s l_s = \frac{1}{4 \pi} \frac{\l}{M_{KK }N } \ , \q \q \q  \frac{R^3}{l_s ^2} = \frac{1}{4} \frac{\l}{M_{KK}}\ .
\label{holomaps}
\ee

It can be shown that, since in the first case $g_{00} (u_T) =0$ and in the second one $g_{00}(u_0) \neq 0$, the high-temperature solution is dual to the deconfined phase and the low-temperature one is dual to the confined phase. By computing the free energy of the two backgrounds, one finds that the system exhibits a first-order phase transition at temperature  $T_c = M_{KK}/2 \pi$.

Let us consider now the fundamental matter sector. In the regime (\ref{regimeofstudy}), the backreaction of the D8/anti-D8-branes can be neglected. As a result, they can be treated in the probe approximation, namely by means of the Dirac-Born-Infeld action for the D8 branes on the original backgrounds. The embedding of the branes on the geometry will then be a solution $x_4=x_4(u)$ of the equation of motion coming from this action and found by asking the D8/anti-D8-brane pair  to be separated by a distance $L$ on the circle $S^1 _{x_4}$, as mentioned above. Let us describe the solutions in the two phases and how they depend on the distance $L$.

In the confined phase,  each D8/anti-D8 branes pair is actually bound to join into a single U-shaped configuration. From the field theory perspective, this fact is interpreted as a realization of chiral symmetry breaking. When $L = \pi M_{KK} ^{-1}$, the branes are antipodal and  join at a value $u_J$ of the holographic coordinate that coincides with the smallest value of the coordinate range, that is $u_J = u_0$. This means that chiral symmetry breaking occurs at the confinement scale. In contrast, when $L < \pi M_{KK} ^{-1}$, the branes join at $u_J>u_0$, meaning that chiral symmetry breaking and confinement can occur at different scales.
In the QCD-like setup, $N_f$ coincident pairs of D8/anti-D8-branes are placed in the antipodal configuration and the model realizes the breaking of $U(N_f)\times U(N_f)$ to the diagonal $U(N_f)$. At low energies, the effective action on the D8-branes reproduces the chiral Lagrangian (with pion decay constant $f_{\pi}\sim \sqrt{N} M_{KK}$) including the Skyrme term and the axial anomaly term that gives mass to the $\eta '$ particle. The model has been generalized in \cite{Bigazzi:2019eks} so that the effective Lagrangian also includes the Peccei-Quinn axion.\footnote{This generalized version of the model has been also used in order to compute the axion coupling to nucleons \cite{Bigazzi:2019hav}.} This can be easily obtained by considering an extra D8/anti-D8 pair placed in a non-antipodal configuration in order to achieve a separation between the axion scale $f_a$ and the QCD confinement scale. The $U(1)$ gauge symmetry of the extra pair of branes is holographically interpreted as the Peccei-Quinn $U(1)_{PQ}$ global symmetry whose breaking gives the axion as a pseudo-Nambu-Goldstone boson.

In the deconfined phase, branes and anti-branes are not bound to join, because they can terminate on the horizon. As a result, depending on $L$, there are two possible D8-brane embeddings. If $L> 0.97 M_{KK}^{-1} $, the branes remain disconnected: the embedding $x_4 = x_4(u)$ reduces to a constant. From the field theory side, this corresponds to chiral symmetry restoration. In contrast, if $L<0.97 M_{KK}  ^{-1}$, both the  connected and the disconnected embeddings are allowed and, depending on temperature, only one is energetically favored. As a result, the model features a further first-order transition, occurring at a critical temperature $T_c ^\chi$ different from the confinement/deconfinement critical temperature $T_c$ and given by $T_c ^\chi \simeq 0.1538/L$. 
For $T > T_c ^\chi$, the disconnected solution is preferred and thus chiral symmetry is restored, while for $T < T_c ^\chi$, the connected one is favored and thus chiral symmetry is broken.

\section{Calculation of the gravitational wave spectra}
\label{appformule}
In this appendix, we review all the formulae needed to calculate the gravitational wave spectra produced by cosmological first-order phase transitions. The formulae for the spectra are reported in section \ref{gwcollision}. They require the knowledge of some crucial parameters which we discuss in section \ref{secparameters}. These are essentially given by the temperature (and hence the related value of the Hubble parameter) at which the phase transition completes, the phase transition duration $\beta^{-1}$, computed starting from the bubble nucleation rate $\Gamma$, the strength $\alpha$, i.e., the energy budget of the transition and the bubble wall speed $v$.

\subsection{Parameters}
\label{secparameters}

\subsubsection{Bubble nucleation rate}
First-order phase transitions are triggered by the nucleation of true vacuum bubbles on the false vacuum state. Such nucleation can occur through thermal or quantum fluctuations. As we will discuss in the following subsections, whether the transition actually takes place depends on the ratio $\G / H^4$, where $\G$ is the bubble nucleation rate per unit of volume and $H$ is the Hubble scale. The latter is determined by the energy density $\r$ through the Friedmann equation $H^2 = \rho/3 M_{Pl}^2$,
where $M_{Pl} \approx 2.4\cdot 10^{18}$ GeV. 

The bubble nucleation rate can be computed using the well-known formalism developed in \cite{Coleman:1977py,Linde:1980tt,Linde:1981zj}
for models where the transition is described by a single field $\Phi$. One has to find a particular solution $\Phi_B$ of the Euclidean  equation of motion usually called \emph{bounce}. The latter satisfies the following boundary conditions: it approaches the false vacuum $\Phi_f$ at Euclidean infinity and a constant $\Phi _0$ at the center of the bubble.\footnote{In \cite{Coleman:1977py,Coleman:1980aw} it is discussed how this Euclidean solution is meant to represent the bubble at time zero in Minkowskian signature.}
When the transition from the false to the true vacuum is due to quantum tunneling, the bounce is $O(4)$ symmetric: in this case $\Phi_B$ only depends on the radial coordinate $\rho=\sqrt{t^2+ x_i x_i}$, where $t$ is the Euclidean time and $x_i$ are the space coordinates. When the transition is (mostly) driven by thermal fluctuations, the bounce is $O(3)$ symmetric: in this case $\Phi_B=\Phi_B(\rho)$, with $\rho=\sqrt{x_i x_i}$.  The configuration which dominates the process is the one for which the rate $\Gamma$ has the larger value. As a result, the formula for the bubble nucleation rate reads
\be 
\label{Gamma2}
\Gamma = {\rm Max}\left[T^4 \left( \frac{S_{3,B}}{2\pi T} \right)^{3/2} e^{-S_{3,B}/T}      ,  \left( \frac{S_{4,B}}{2\pi \rho_w^2}  \right)^2  e^{-S_{4,B}}   \right]\ ,
\ee
where $\rho_w$ is the size of the $O(4)$ bubble. The bounce action $S_{3,B}$ appearing in (\ref{Gamma2}) is defined by $S_{3,B}/T  = (S_3(\Phi_B) - S_3(\Phi_f))/T$, where $S_3(\Phi)$ is the $O(3)$-symmetric Euclidean action for the scalar field. The action $S_{4,B}$ is defined analogously.


\subsubsection{The relevant temperatures}
\label{relevanttemperatures}
In order to calculate the spectrum of GWs, the first datum to determine is the temperature at which the waves are produced.
Since from the time of nucleation, which happens at plasma temperature $T_n$,  to the time where most of the collisions take place and most of the sound waves collide there could be a sizable difference, the percolation temperature $T_p$ is considered to be the relevant one for the production of gravitational waves \cite{Caprini:2019egz}. In the following, we are going to discuss both $T_n$ and $T_p$.

\subsubsection*{Nucleation temperature}
The nucleation time $t_n$ is defined as the time at which the total number of nucleated bubbles per Hubble patch from $t=t_c$ (the time when the Universe is at the critical temperature $T_c$) to $t=t_n$ is order one,
\be
\label{conditiontnt}
\int_{t_c}^{t_n}dt \frac{\Gamma}{H^3} \sim 1\ ,
\ee
where $H = \dot R (t)/R(t)$ is the Hubble scale. We can write this condition in terms of the temperature of the Universe.
Assuming\footnote{When the energy density behaves as $\r \sim g_* T^4$ with a time-independent number of relativistic  degrees of freedom $g_*$, $r=1$. In general, $g_*$ may depend on the temperature. In the WSS model, in the regime where the contribution from the glue sector dominates, $r = 5/3$.}  $R(T) \sim T^{-r}$, we have 
\be
\label{timetemperature}
r \frac{dT}{T} =- H d t \ ,
\ee
and therefore (\ref{conditiontnt}) becomes
\be
\label{conditiontntot}
r \int_{T_n}^{T_c}\frac{dT}{T} \frac{\Gamma}{H^4} \sim 1\ .
\ee
We can get analytical insight by noticing that the integral is dominated by the region very close to $T_n$. 
The general form of the nucleation rate is
\be
\label{formnucleationrate}
\Gamma (T) = f(T)  {\rm \exp}(-S_B(T) )\ ,
\ee
where $f(T)$ is a polynomial function, usually assumed to be $T^4$ from dimensional analysis.
Let us write the Taylor expansion of the exponent as
\be 
\label{taylorexpansionaction}
S_B(T) \sim S_B(T_n) + (T-T_n) \frac{\tilde\beta r}{H T}|_{T_n}\ ,
\ee
where
\be
\tilde \beta \equiv - \frac{d S_B}{d t} = \frac{H T}{r} \frac{dS_B}{d T}\ .
\label{tildebeta}
\ee
Thus, the condition (\ref{conditiontntot}) can be approximately computed as
\be
1 \sim r \pr{ \frac{\Gamma }{H^4 T} } |_{T_n} \int_{T_n}^{\infty} dT  \, e^{- \pr{ \frac{\tilde\beta r}{H T}}|_{T_n} (T-T_n)} \ ,
\ee
where we extended the integration domain to infinity, and therefore it reads
\be\label{conditiontnI} 
\frac{\Gamma}{H^4}|_{T_n} \sim \frac{\tilde\beta}{H}|_{T_n}\ .
\ee

\subsubsection*{Percolation temperature}

The percolation temperature $T_p$ is defined as the Universe temperature when the fraction of space sitting in the true vacuum takes a benchmark conventional value. We choose the latter to be one.\footnote{Another value that is often taken in the literature is 0.34. We have verified that in our cases the gravitational wave spectra are not significantly sensitive to such a difference.}
In order to compute the percolation temperature, we have to estimate the size of a bubble as a function of time, which involves the knowledge of the bubble wall speed $v$. We follow \cite{Ellis:2018mja}. The fraction of space in the true vacuum reads
\be
\label{fractiontruevacuum}
I(t) = \frac{4 \pi}{3} \int _{t_c} ^t dt' \G (t') R(t') ^3 r_b(t,t') ^3 \ ,
\ee
where $r_b(t,t')$ is the size of the bubble in comoving coordinates as a function of time, which can be obtained by
\be
r_b(t,t') = \int _{t'} ^t \frac{d \tilde t \, v}{R(\tilde t)} \ .
\ee
Here, $v$ is the velocity of the bubble wall. Using $R(T) \sim  T^{-r}$ and (\ref{timetemperature}), we have
\be
\label{exactfraction}
I(T) = \frac{4 \pi r^4}{3} \int _{T} ^{T_c} \frac{dT' \G (T')}{H(T') T'^{1+3r}} \pr{ \int _{T} ^{T'} \frac{d \tilde T \, v}{H(\tilde T) \tilde T ^{1-r}}}^3 \ .
\ee
We therefore define the percolation temperature $T_p$ by
\be\label{condit}
I(T_p) = 1 \ .
\ee

In the scenarios considered in this paper, the energy density includes a radiation term and a vacuum term. Adopting the notation of \cite{Ellis:2018mja}, we therefore write $H= H_R + H_V$. Let us consider approximate solutions to (\ref{exactfraction}). Firstly, let us consider the case in which the vacuum contribution $H_V$ can be neglected. This is expected to give a good approximation when supercooling is not significant. Assuming $H_R = c_{R} T^s$, and constant velocity $v$, we obtain
\be
\label{fractionradiation}
I(T) = \frac{4 \pi r^4 v^3}{3 c_{R} ^4 (s-r)^3} \int _{T} ^{T_c} \frac{dT' \G (T')}{ T'^{1+3r+s}} \pr{\frac{1}{T^{s-r}} - \frac{1}{T'^{s-r}}}^3 \ .
\ee
The formulae of reference \cite{Ellis:2018mja} are retrieved putting $s=2$, $r=1$, and $c_R = (\sqrt{3} M_{Pl} \xi _g)^{-1}$. In the WSS model (see (\ref{energydarkgauge})), 
\be
s=3 \ , \q \q r=5/3 \ , \q \q  c_{R} = \sqrt{5} \frac{2^3  \pi ^2}{3^4} \frac{\sqrt{g}}{M_{Pl} M_{KK}} \ . 
\ee
Evaluating the integral (\ref{fractionradiation}) as done above for the nucleation temperature, we find an approximate formula for the percolation temperature $T_p$, which does not depend on the coefficients $r$ and $s$,
\be
\label{percolationtemperatureradiationnew}
\frac{\Gamma}{H^4}|_{T_p} \approx \frac{1}{8 \pi v^3}\left(\frac{\tilde\beta}{H}\right)^4|_{T_p} \ .
\ee

When there is supercooling, the vacuum term $H_V$ may become dominant before percolation. Defining the temperature $T_V$ by $H_R(T_V) = H_V(T_V)$, let us approximate the Hubble scale with
\be
H(T) = H_R \Theta (T-T_V) + H_V \Theta (-T+T_V) \ ,
\ee
where $\Theta (\cdot)$ is the Heaviside step function. In this case, the factor $R (T') r_b (T,T')$ appearing in (\ref{fractiontruevacuum}) takes two contributions, reading
\be
R (T') r_b (T,T') = \frac{v}{H_V} \pq{ 1- \pr{\frac{T}{T'}}^r}
\ee
for $T \leq T' \leq T_V$, and
\be
R (T') r_b (T,T') = \frac{v}{H_V} \pq{ \frac{s}{s-r} \pr{\frac{T_V}{T'} }^r - \frac{r}{s-r} \pr{\frac{T_V}{T'}}^s -  \pr{\frac{T}{T'}}^r}
\ee
for $T \leq T_V \leq T '$. As a result, the fraction of volume in the true vacuum takes the form,
\ba
\label{percolationtemperaturevacuumnew}
I(T) &=& \frac{4 \pi r v^3}{3 H_V ^4} \Bigg\{ \int_{T_V}^{T_c} \frac{dT' \Gamma(T')}{T'^{1+s}} T_V ^{s} \pq{ \frac{s}{s-r} \pr{\frac{T_V}{T'} }^r - \frac{r}{s-r} \pr{\frac{T_V}{T'}}^s -  \pr{\frac{T}{T'}}^r}^3 + \nb \\
&+&  \int_{T}^{T_V} \frac{dT' \Gamma(T')}{T'} \pq{ 1- \pr{\frac{T}{T'}}^r}^3   \Bigg\} \ . 
\ea
Notice that if $T_p=T_V$, $I(T_p)$ from (\ref{percolationtemperaturevacuumnew}) precisely reduces to the value computed using (\ref{fractionradiation}). Hence, in general, when $T_p\approx T_V$ we can still use formula (\ref{percolationtemperatureradiationnew}) to estimate the percolation temperature. The same conclusion holds in the different limit $T_p\ll T_V\approx T_c$. In all the cases examined in this paper we have found no notable numerical differences between the percolation temperature computed using formula (\ref{fractionradiation}) and that computed using (\ref{percolationtemperaturevacuumnew}). 

Finally, when the Universe is inflating due to vacuum energy domination, it is not guaranteed that the transition can complete at all, since the bubbles can never percolate with the required velocity. One needs to check explicitly that the probability of finding a fraction of space occupied by the false vacuum, $V_{false} \propto R(t)^3 \exp \pr{-I_{RV}}$, is decreasing at the supposed percolation temperature.
This translates into the condition
\be
\label{transitionok}
\frac{1}{V_{false}}\frac{dV_{false}}{dt} = H(T_p)\left(3 + \frac{T_p}{r} \frac{dI_{RV}(T)}{dT}|_{T_p}\right) < 0.
\ee

Once the percolation temperature has been determined, one can derive a crucial parameter for the spectrum
\be
\label{minusfour}
\frac{\beta}{H_{*}} = \frac{1}{H_* \G} \frac{d \G}{dt}|_{T_p} =  - \frac{1}{\Gamma} \frac{T}{r} \frac{d\Gamma}{dT}|_{T_p}\,.
\ee

\subsubsection*{Reheating temperature}
\label{reheat}
During a first-order phase transition, entropy is released and therefore the Universe gets heated. Assuming that the entropy release is approximately instantaneous, we define the reheating temperature $T_R$ as the temperature of the Universe after the release. By exploiting the conservation of the energy density during the transition, we find the reheating temperature through the formula
\be
\label{condtR}
\rho_t(T_R) = \rho_f(T_p)\,,
\ee
where $\rho_f$ and $\rho_t$ are, respectively, the total energy density in the false and true vacua.

Especially in the case of strong first order transitions, like the ones examined in this paper, the reheating temperature may be greater than the critical temperature $T_c$ (see e.g. \cite{Nardini:2007me}). In this case, one should check whether the inverse phase transition could take place or not. In the cases examined in this paper this does not happen essentially because the distance in field space between the two minima of the effective potential at $T=T_R\gg T_c$ is ``large" enough to drastically suppress the rate of the inverse transition w.r.t. the Hubble scale.

\subsubsection{Released energy and wall velocity}
Another crucial parameter for the gravitational wave spectra is the ratio of the energy released in the transition to the energy  of the radiation bath \cite{Caprini:2019egz}. In particular, the formulae for the spectra include the parameter $\a$ defined as
\be\label{alphacap}
\alpha = \frac{\Delta\theta}{\rho_{rad}}\ ,
\ee
where $\theta= (\rho-3p)/4$ is the trace of the energy-momentum tensor, and the $\Delta$ indicates the difference between the false and true vacua.

The knowledge of the parameter $\alpha$ allows us to estimate the velocity of the bubble walls according to the Chapman-Jouguet formula
\be
\label{vJC}
v = \frac{1/\sqrt{3}+\sqrt{\alpha^2 + 2\alpha/3}}{1+\alpha}\ .
\ee
Formula (\ref{vJC}) has a limited range of validity; in particular, it has to be corrected when the friction in the bubble interactions with the plasma is significant. However, to provide better estimates of the wall speed is still one of the big open problems in determining the bubble dynamics.

In the case in which we consider a dark sector that is not in thermal equilibrium with the visible one, we have to define two separated $\a$ parameters for the two sectors,
\be 
\label{twoalphas}
\alpha = \frac{\Delta\theta}{\rho_{rad,SM}}\ , \q \q \quad \alpha_D = \frac{\Delta\theta}{\rho_{rad,glue}}\ ,
\ee
which take into account the fact that the relevant radiation could be only the one of the visible (dark) sector, $\rho_{rad,SM}$ ($\rho_{rad,glue}$).
Note that if the Standard Model plasma is not interacting with the dark sector one, in formula (\ref{vJC}) one has to replace $\a$ with $\alpha_D$.

\subsubsection{Redshift}
\label{sec:redshift}
Once we know the parameters that we have discussed so far, we can compute the spectrum of gravitational waves as it appears at the time of production. From this time to the time of detection, the signal gets redshifted due to the cosmological expansion. We are going to discuss how to take into account the redshift of the signal in two different circumstances, both occurring in the scenarios that we study in the present paper.

Let us start with the case in which the Universe evolves adiabatically from gravitational waves emission to the detection time \cite{Kamionkowski:1993fg}. This is the case in which only one first-order phase transition occurs. Hence, it includes all the scenarios that we consider in this paper but the Dark HQCD 2 one. Let us call $T_e$ and $T_d$ the temperature of the Universe, respectively, at the emission and at the detection times. 
The detection temperature is $T_d \sim 2.35 \cdot 10^{-13}$ GeV. The adiabatic evolution is characterized by the conservation of the entropy
\be
S \sim R^3 g_*^S(T) T^3\ ,
\ee
from which we find the ratio of the scale factors between the two temperatures
\be
\label{scalefactors}
\frac{R_d}{R_e}= \left(\frac{g_{*,e}^S}{g_{*,d}^S} \right)^{1/3} \frac{T_e}{T_d}\ .
\ee
In this expression, $g_{*,e}^S $ and $g_{*,d}^S$ are the number of relativistic degrees of freedom at the time of emission and detection, respectively; they are computed in the free case using the general formula 
\be
g_* ^S (T) = \sum_{i = \text{bosons}} g_i \pr{\frac{T_i}{T}}^3 + \frac{7}{8} \sum_{i = \text{fermions}} g_i \pr{\frac{T_i}{T}}^3  \ ,
\ee
where $T_i$ represents the temperature of the $i$-th species.

The frequency $f$ and the energy density\footnote{As customary in cosmology, $\O$ is defined as the energy density divided by the critical density $\r_{crit,0} = 3 M_{Pl} ^2 H_0 ^2$, where $H_0$ is the Hubble scale computed in the present epoch.} $\Omega$ of the GWs get redshifted as $R^{-1}$ and $R^{-4}$ respectively, hence
\begin{subequations}
\ba
 f_d   &=& f_e \frac{R_e}{R_d}\ , \\
 H_d^2 \Omega_d &=&  H_e^2 \Omega_e \left( \frac{R_e}{R_d} \right)^4\ .
\ea
\end{subequations}
The Hubble scale $H$ is given by the energy density via the Friedmann equation
\be
\label{hubblehere}
H^2 = \frac{\r}{3 M_{Pl}^2} = \frac{1}{3 M_{Pl}^2} \frac{\pi^2}{30} g_*(T) T^4 \ .
\ee
Here, $g_*$ is defined in the free case as\footnote{Notice that if some species are decoupled from the bath, $g_* \neq g_* ^S$. This, notoriously, occurs in the cosmological evolution because of neutrino decoupling when electrons and positrons become non-relativistic. Neutrino and photon relic temperatures do not coincide. They are related by $T_\n = (4/11)^{1/3} T_\g$. As a result, today $g_*  \approx 3.36$ is different from $g_* ^S = 3.91$.}
\be
g_* (T) = \sum_{i = \text{bosons}} g_i \pr{\frac{T_i}{T}}^4 + \frac{7}{8} \sum_{i = \text{fermions}} g_i \pr{\frac{T_i}{T}}^4  \ .
\ee
As a result, using (\ref{scalefactors}) and (\ref{hubblehere}), we find
\begin{subequations}
\label{redshift12}
\ba
\label{redshift1}
 f_d &=& \frac{\pi}{\sqrt{90 M_{Pl}^2}} \frac{f_e}{H_e}T_e T_d \frac{(g_{*,d}^S)^{1/3}(g_{*,e})^{1/2}}{(g_{*,e}^S)^{1/3}}\ , \\
 H_d^2 \Omega_d &=& \frac{\pi^2}{90 M_{Pl}^2}  \Omega_e T_d^4 \frac{(g_{*,d}^S)^{4/3}(g_{*,e})}{(g_{*,e}^S)^{4/3}}\ . 
 \label{redshift2}
\ea
\end{subequations}

Let us now consider the case in which two first-order phase transitions occur. Among the scenarios studied in this paper, this happens in the Dark HQCD 2 scenario of section \ref{secdarkqcd2}, where a chiral symmetry breaking/restoration transition
is followed by a confinement/deconfinement one. We will refer to this case, even though the discussion will be valid for two generic separated first-order phase transitions.

When we compute the redshift of the gravitational waves spectrum associated with the chiral symmetry transition, we have to take into account that adiabaticity is violated during the confinement/deconfinement one. As a result, conservation of entropy can be used from the time of the chiral symmetry breaking transition, where the temperature $T_e$ is taken to be the reheating temperature, to the percolation time of the confinement transition $T_{p,conf}$.  Then, assuming fast reheating in the confinement transition, the temperature has a sudden jump from the percolation temperature $T_{p,conf}$ to the reheating temperature $T_{R,conf}$. Finally, from this time to the present,  the Universe evolves adiabatically, and we can again use the conservation of entropy. All in all, the redshifted frequency and energy density read
\ba
\label{fd2}
f_d &=& f_{e} \frac{R_e}{R_{p,conf}} \frac{R_{R,conf}}{R_d} = f_{e} \left( \frac{g_{*,p,conf}^S}{g_{*,e}^S} \right)^{1/3} \frac{T_{p,conf}}{T_e}\left( \frac{g_{*,d}^S}{g_{*,R,conf}^S} \right)^{1/3} \frac{T_{d}}{T_{R,conf}} \nonumber \\
&& =\frac{\pi}{\sqrt{90 M_{Pl}^2}} \frac{f_e}{H_e}T_e T_d \frac{(g_{*,d}^S)^{1/3}(g_{*,e})^{1/2}}{(g_{*,e}^S)^{1/3}} \cdot \frac{(g_{*,p,conf}^S)^{1/3} T_{p,conf}}{(g_{*,R,conf}^S)^{1/3}T_{R,conf}}\ .
\ea
and
\be\label{Hd2}
H_d^2 \Omega_d = \frac{\pi^2}{90 M_{Pl}^2}  \Omega_e T_d^4 \frac{(g_{*,d}^S)^{4/3}(g_{*,e})}{(g_{*,e}^S)^{4/3}}  \cdot \left(\frac{(g_{*,p,conf}^S)^{1/3} T_{p,conf}}{(g_{*,R,conf}^S)^{1/3}T_{R,conf}}\right)^4 \ . 
\ee
With respect to the single-transition case, the difference is encoded in the parameter
\be
\label{delta}
\delta \equiv \frac{(g_{*,p,conf}^S)^{1/3} T_{p,conf}}{(g_{*,R,conf}^S)^{1/3}T_{R,conf}} \ .
\ee
In models with multiple, separated phase transitions, a $\delta$  factor for each transition after the first one must be included in the formulae for the GW spectra.

\subsection{Formulae for the spectra}
\label{gwcollision}

Let us finally discuss the formulae that allow us to find the gravitational wave spectrum. In linear approximation, the spectrum is given by the sum of three contributions, coming from the collisions of the bubbles, from collisions of plasma sound waves and plasma turbulence,
\be
h^2 \Omega_{GW}  \approx h^2 \Omega_c (f) + h^2\Omega_{sw} + h^2 \Omega_{turb} \ .
\ee
Here, $h$ is defined from today's value of the Hubble scale through $H_0=100\, h \text{Km/s/Mpc}$. 
Following \cite{Caprini:2019egz}, we are going to neglect the turbulence contribution because it is still not well-understood and because it is expected that only a small fraction of the transition energy is converted to turbulence.
 
Let us first consider the collision contribution. Using the so-called \emph{envelope approximation}, a formula for the signal of gravitational waves coming from bubble collisions was numerically found in \cite{Kamionkowski:1993fg}. An improved version of such a formula (see, e.g., the review \cite{Weir:2017wfa}) reads
\begin{subequations}
\label{formulascollision}
\be
\label{spectrum}
h^2 \Omega_c(f) \sim 1.67 \cdot 10^{-5} \left(\frac{\beta}{H _*} \right)^{-2} \left(\frac{\kappa \alpha}{1+\alpha} \right)^2 \left(\frac{100}{g_*} \right)^{1/3} \left(\frac{0.48 v^3}{1+5.3v^2+5v^4} \right) S_{env}(f)
\ , 
\ee
where $f$ denotes the frequency of the waves, and $v$ the average velocity of the bubbles. The factor $\kappa$ quantifies the fraction of available energy converted into gravitational waves coming from bubble collision. Finally, the spectral form $S_{env}$ and the peak frequency $f_{env}$ are given by
\ba
S_{env}(f) &\sim& \left[0.064 \left(\frac{f}{f_{env}} \right)^{-3} +    0.456 \left(\frac{f}{f_{env}} \right)^{-1} + 0.48 \left(\frac{f}{f_{env}} \right)  \right] \ , \\
\label{peak}
 f_{env} &\sim& 16.5 \cdot 10^{-6} {\rm Hz} \left(\frac{f_*}{\beta}\right) \left(\frac{\beta}{H_*} \right) \left(\frac{T_*}{100 {\rm GeV}} \right)\left(\frac{g_*}{100} \right)^{1/6}\ ,
\ea
where
\be
\frac{f_*}{\beta} \sim \frac{0.35}{1+0.069 v +0.69 v^4}\ .
\ee
\end{subequations}
In these formulae, $\b/H_*$, $g_*$, and $\a$ are evaluated at the percolation temperature $T_p$, whereas $T_*$ in (\ref{peak}) is identified with the reheating temperature.\footnote{If supercooling is small, reheating is small as well, and therefore the reheating and nucleation temperatures approximately coincide. This is why, in the literature, $T_*$ is often taken to be the nucleation temperature.}

Until recently, the sound wave contribution $\Omega_{sw}$ was expected to be subleading with respect to $\Omega_c$ in the $v \sim 1$ limiting case. Indeed, scenarios with $v \sim 1$ are expected to be characterized by large supercooling, which causes the plasma to be very diluted, and friction effects to be suppressed. Such a scenario was challenged in \cite{Bodeker:2017cim}, where it was pointed out that even in these conditions, there is a so-called \emph{transition radiation} given by the emission of particles that change mass across the bubble walls, which cause a friction pressure.\footnote{See also \cite{Baldes:2020kam} for a recent discussion in the context of confining phase transitions.} This friction causes (at least part of) the energy to be transmitted to the plasma rather than stored in the bubble wall kinetic energy. As a result, whether $\Omega_c$ or $\Omega_{sw}$ dominates the spectrum depends on the energy fraction that gets dispersed in the plasma. This is a highly non-trivial quantity to calculate.

The spectrum due to sound waves is given by\footnote{A word of caution is in order. The known formulae for $\Omega_{sw}$ have been derived under the hypothesis that $\alpha \lesssim 0.1$ and that the speed is far from the Chapman-Jouguet one. 
Lacking better estimates, these formulae are usually employed even when $\alpha$ is larger than this value. A first study of the spectrum for $\alpha \sim 1$ has highlighted a further suppression of the signal \cite{Cutting:2019zws}. Nevertheless, this suppression is more important in the case of deflagration, which is not relevant in our cases.} \cite{Hindmarsh:2017gnf,Weir:2017wfa}
\begin{subequations}
\label{formulasoundwaves}
\be
\label{spectrumOsw}
h^2 \Omega_{sw}(f) \sim 8.5 \cdot 10^{-6} \left(\frac{\beta}{H_*} \right)^{-1} \left(\frac{\kappa_v \alpha}{1+\alpha} \right)^2 \left(\frac{100}{g_*} \right)^{1/3} v\, S_{sw}(f) \ .
\ee
The spectral shape and peak frequency today in this case are
\ba
\label{peaksw}
 S_{sw}(f) &\sim & \left(\frac{f}{f_{sw}}\right)^{3}\left(\frac{7 }{4+3(f/f_{sw})^{2}} \right)^{7/2}\ , \\
 f_{sw}    &\sim & 8.9 \cdot 10^{-6} {\rm Hz} \frac{1}{v} \left(\frac{\beta}{H _*} \right) \left(\frac{z_p}{10} \right)\left(\frac{T_*}{100 {\rm GeV}} \right)\left(\frac{g_*}{100} \right)^{1/6}\ ,
\ea
\end{subequations}
where we are going to use the approximate value $z_p \sim 10$, and the efficiency factor in the case $v \sim 1$ is
\be\label{kappav}
\kappa_v = \frac{\alpha}{0.73+0.083\sqrt{\alpha}+\alpha}\ .
\ee
If the Standard Model plasma is not in thermal equilibrium with the dark sector, in this formula one has to use $\alpha_D$ instead of $\alpha$ \cite{Fairbairn:2019xog}. 

In fact, formula (\ref{spectrumOsw})  
is valid under the assumption that the source of GWs lasts for a period longer than a Hubble time.
If the source's duration is short, turbulence effects can be sizable and one can estimate that the net effect is to multiply formula (\ref{spectrumOsw}) by a factor \cite{Caprini:2019egz}
\be \label{factor}
(8 \pi)^{1/3} v \left(\frac{\beta}{H_*} \right)^{-1} \left(\frac{\kappa_v \alpha}{1+\alpha}  \right)^{-1/2}\ .
\ee
This term tends to reduce the amplitude of the signal.
On the other hand, one should add the contribution due to turbulence, which is very uncertain and as stated it is ignored in this paper.
Thus, $\Omega_{sw}$ including the term (\ref{factor}) really corresponds to a lower bound on the contribution of the plasma to the GW spectrum. 
See also \cite{Guo:2020grp} for a discussion of this topic.


\section{Holographic bubbles}
\label{sec:holobubbles}

In this appendix, we review the results of \cite{Bigazzi:2020phm} that are used in order to compute the spectrum of gravitational waves. In \cite{Bigazzi:2020phm}, the confinement/deconfinement phase transition occurring in the WSS model was studied using an effective approach inspired by \cite{Creminelli:2001th}, and deployed in order to reduce the problem to a single-scalar one (a recent alternative approximation can be found in \cite{Agashe:2020lfz}).
Moreover, the chiral symmetry breaking/restoration transition was described by the deformation of the D8 embedding, again encoded in a single scalar mode.
We start with the case of the confinement/deconfinement phase transition and then we discuss the chiral symmetry breaking/restoration one.

\subsection{Confinement/deconfinement phase transition}
The main idea for studying the confinement/deconfinement phase transition is to promote the parameters $u_0$ and $u_T$ in (\ref{wittenbackground}) and (\ref{solitonicW}) to fields depending on a radial coordinate $\rho$ in such a way that the background displays a conical singularity. For the case in which the transition is driven by thermal fluctuations and the bounce is $O(3)$ symmetric, this can be achieved by taking the ansatz
\be
\label{secondansatz}
ds^2 = \pr{\frac{u}{R}}^{3/2} \pq{f_T(u) dt^2 + d\rho^2 +\rho^2 d\Omega_2^2 +  d x_4 ^2 } +\pr{\frac{R}{u}}^{3/2} \pq{\frac{9\,u_T\, r^2 dr^2}{4 R^3 f_T(u)} +u^2 
d \O _4 ^2} \ ,
\ee
where 
\be
u = u(r,\rho) = u_T(\rho) + \frac{3}{4} \sqrt{\frac{u_T(\rho)}{R^3}} r^2
\label{redef1}
\ee
and
\be
\label{utthrho}
f_T(u,\r) =1- \frac{u_{T} (\r) ^3}{u^3}\ , \q \q u_T(\r) = \frac{16 \pi ^2}{9} R^3\, T_h (\r) ^2 \ .
\ee
The other fields are left unchanged. An analogous ansatz holds for the  background dual to the confined phase. By computing the free energy through the holographic renormalization procedure, we find the effective action from which we can compute the bounce solution and the bubble nucleation rate. In terms of the dimensionless field
\be
\Phi (\r) =
-\frac{T_h^2 (\r)}{M_{KK} ^2} \q \q \text{for $\Phi < 0$}  \ , \q \q 
\Phi (\r) = \frac{M_h^2 (\r)}{M_{KK} ^2} \q \q \text{for $\Phi > 0$} \ ,
\ee
and of the dimensionless quantities
\be
\label{dimsq}
\bar \rho \equiv M_{KK} \rho\ , \q \q \quad \bar T \equiv \frac{2\pi T}{M_{KK}}\ , 
\ee
the effective action reads
\be
\frac{S_3 (\Phi)}{T}=\frac{32\pi^4 g}{3^5 \bar T} \int_{0}^{\infty}d\bar\rho \bar\rho^2 \left [ \left(5-\frac{\pi}{2\sqrt{3}} \right) \Phi'^2 +  \Theta(\Phi) V_c(\Phi) + \Theta(-\Phi) V_d(\Phi) \right]\,,
\label{o3dc}
\ee
where
\be
g \equiv \lambda N^2\,,
\ee
$\Theta(\cdot)$ is the Heaviside step function and
\bea
V_c(\Phi) &=& \frac{16\pi^2}{9}\left(5\Phi^3-\frac{3}{\pi}\Phi^{5/2}\right)\,,\nonumber \\
V_d(\Phi) &=& -\frac{16\pi^2}{9}\left(5\Phi^3+\frac{3}{\pi}\bar T (-\Phi)^{5/2}\right)\,.
\eea

\begin{figure}
\center
\includegraphics[scale=0.5]{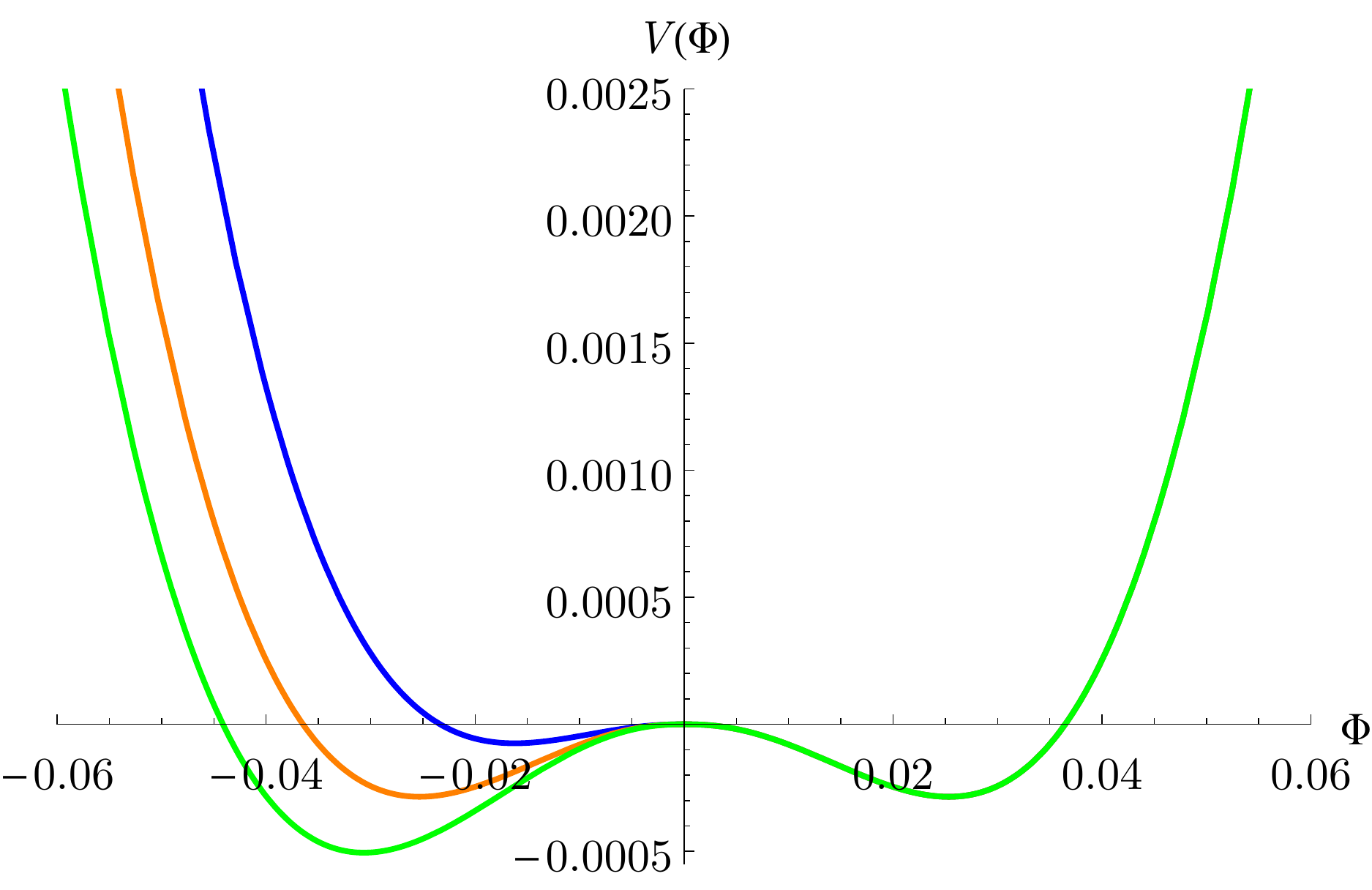}
\caption{Plots of the potential for $\bar T=0.8$ (blue), $\bar T=1$ (orange), $\bar T=1.1$ (green). The region where $\Phi$ takes positive values does not depend on the temperature, hence the curves overlap.}
 \label{figpot}
\end{figure}

The shape of the full potential is shown in figure \ref{figpot} for three different values of the reduced temperature $\bar T$. The two minima are $V_d=-\bar T^6/(36 \pi^4)$ for $\Phi_d=-\bar T^2/(4\pi^2)$ and $V_c=-1/(36 \pi^4)$ for $\Phi_c= 1/(4\pi^2)$. In the following we will focus on the case $\bar T \in [0,1]$, where the true vacuum is the confining one at $\Phi=\Phi_c$.

The bounce solution $\Phi _B$ is found solving the equation of motion following from the action (\ref{o3dc}) with boundary conditions
\be
\Phi' _B(0)=0  \ , \q \q \q \lim_{\r \to\infty} \Phi _B(\r) = \Phi _d \ .
\ee
Once the solution is found, one can plug it back into the action. As we have already pointed out, what really enters the formula for the nucleation rate is the difference between the on-shell action on the bounce solution and the action evaluated on the false vacuum,
\be
\frac{S_{3,B}}{T} = \frac{S_3(\Phi_B) - S_3(\Phi_d)}{T}  \ .
\ee

From the numerical results and the functional form of the thin and thick wall approximations studied in Appendix A of \cite{Bigazzi:2020phm},
a continuous analytic approximation to the action for the $O(3)$ bubble can be provided as follows, 
\begin{equation}
\frac{S_{3,B} }{g T} \approx \begin{cases}
0.32\ \bar{T}^{5/2} \qquad\qquad\qquad\qquad\qquad\qquad\ (\bar{T} \leq 0.3)\\
1.8 \cdot 10^{-3} \exp (7.9\ \bar T) - 2\times 10^{-3} \qquad ( 0.3\leq \bar T  \leq 0.68)\\
5.4 \cdot 10^{-2} \exp (8.8\ \bar T^{3.8}) \qquad\qquad \qquad   ( 0.68\leq \bar T  \leq 0.87)\\
2.6/\bar T(1\, -\bar T^6)^2\qquad \qquad\qquad\qquad \quad \, \,  \, (\bar T \geq 0.87)
\end{cases}
\label{fitsO3}
\end{equation}

For small temperatures, one could also have $O(4)$ symmetric bounces.
The action for this case is simply (\ref{o3dc}) generalized such that it enjoys $O(4)$-symmetry,
\be\label{o4dc}
S_4 (\Phi) = \frac{8\pi^4 g}{3^5} \int_{0}^{\infty}d\bar\rho\,\bar\rho^3 \left[ \left(5-\frac{\pi}{2\sqrt{3}} \right) \Phi'^2 + \Theta(\Phi) V_c(\Phi) + \Theta(-\Phi) V_d(\Phi) \right]\,.
\ee

From this action, we find the bounce solution imposing the same boundary conditions as above.
For the $O(4)$ bubble, since it is only defined for small temperatures, it is sufficient to consider 
the functional form of the thick wall approximation, giving
\begin{equation}
\frac{S_{4,B} }{g} \approx 0.39\ \bar{T}^{3} \ , \qquad \qquad \bar \rho_w \approx \frac{4.0}{\bar T^{1/2}}  \qquad\qquad\ (\bar{T} < 0.06) \ .\label{fitsO4}
\end{equation}
Let us recall that the $O(4)$ configuration can be admitted only if the bubble radius is much smaller than $1/T$, \cite{Linde:1980tt,Linde:1981zj}. 
 In \cite{Bigazzi:2020phm} the following convention has been adopted: the maximal allowed radius for the $O(4)$ configuration to be considered is set by $\rho_w = 1/2\pi T$ (the radius of the thermal circle).
In the present setup, this corresponds to $\bar T \approx 0.06$, which explains the condition in parenthesis in (\ref{fitsO4}).

\subsection{Chiral symmetry breaking/restoration phase transition}
As we have reviewed in section \ref{basicsecWSSreview} and appendix  \ref{secWSSreviewfeatures}, in the regime 
\begin{equation}
\frac{M_{KK}}{2\pi} < \frac{0.1538}{L}
\end{equation}
the WSS model displays a first-order phase transition associated with chiral symmetry breaking in the deconfined phase.
In the probe regime $N_f \ll N$, the phase transition can be studied just by considering the Dirac-Born-Infeld action on the fixed black hole background (\ref{wittenbackground}) which, using spherical coordinates for the 3d Euclidean physical space reads
\begin{equation}
ds_E^2= \left(\frac{u}{R}\right)^{3/2} \left[f_T(u) dt^2 +d\rho^2 + \rho^2 d\Omega_2^2 + dx_4^2\right] + \left(\frac{R}{u}\right)^{3/2} \left[\frac{du^2}{f_T(u) }+u^2 d\Omega_4^2 \right] \ .
\label{deconf_metric2}
\end{equation}
Considering an ansatz in which the embedding of the D8-branes in the above background is described by a function $x_4(u,\rho)$, the DBI action reads
\begin{equation}
S_{DBI}=\frac{T_8}{g_s}\int d^9x \rho^2 \left(\frac{u}{R}\right)^{-3/2}u^4 \sqrt{1 + f_T(u) \left(\frac{u}{R}\right)^3 (\partial_u x_4)^2 +(\partial_\rho x_4)^2}  \ ,
\label{SDBI3}
\end{equation}
where $T_8$ is the D8-brane tension.
Let us now rescale the coordinates as follows. First, define
\begin{equation}
x_4 = x \,  u_T^{-1/2} R^{3/2} = x \frac{3}{4\pi T} \ , \qquad u = y\,u_T   \ ,  \qquad u_J = y_J\,u_T\,,
\label{redef}
\end{equation}
so that
the periodicity of the cigar coordinate now reads
\begin{equation}
x \sim x +  \frac{2\pi\sqrt{ u_T}}{M_{KK} R^\frac32}=
x +  \frac{8\pi^2 T}{3 M_{KK}}\,,
\end{equation}
and
\begin{equation}
f_T(u) \equiv f_T = 1 - y^{-3}  \ , \qquad  f_T(u_J)  \equiv f_{TJ} = 1 - y_J^{-3} \ .
\label{redef2}
\end{equation}
Then, let us define 
\begin{equation}
\rho = \s \,  u_T^{-1/2} R^{3/2} = \s \frac{3}{4\pi T} \,.
\end{equation}
Using the definitions above, the DBI action can be rewritten as
\be
S_{DBI}=\frac{N T^3 \lambda^3}{486 M_{KK}^3}\ \tilde S \ ,
\label{factors_action}
\ee
where
\begin{equation}
\tilde S =\int \int \s ^2  y^{5/2} \sqrt{1 + (y^3-1) (\partial_y x)^2 + (\partial_\s x)^2}    d \s dy \ .
\label{generalS2}
\end{equation}
Once extracted the factor written in (\ref{factors_action}), the renormalized on-shell action is
\be
\label{vari_acti2}
\Delta \tilde S = 2 \int_0^\infty d \s \, \s^2 \left(  \int_{y_J(\s)}^\infty  y^{5/2} \left[ \sqrt{1 + (y^3-1)  (\partial_y x)^2+ (\partial_\s x)^2} -1\right]  dy-  \frac27 (y_J(\s)^{7/2} -1)
\right) \ ,
\ee
where we have subtracted the contribution of the straight brane/antibrane pair configuration. The Euler-Lagrange equation for $x(y,\s)$ reads
\begin{equation}
\partial_y\left(\frac{\s ^2 y^{5/2}  (y^3-1) (\partial_y x)}{ \sqrt{1 + (y^3-1) (\partial_y x)^2 + (\partial_\s x)^2}}
\right)+
\partial_\s \left( \frac{\s^2 y^{5/2}  (\partial_\s x)}{ \sqrt{1 + (y^3-1) (\partial_y x)^2 + (\partial_\s x)^2}}
\right)=0 \ .
\label{fullPDE}
\end{equation}
This is a non-linear partial differential equation, which is extremely hard to solve even numerically. An escape strategy has then been put forward in \cite{Bigazzi:2020phm}: it amounts to look for approximate solutions by using a reasonable variational ansatz,
\begin{equation}
x=\frac{\tilde L}{2} \tanh\left(\frac{\sqrt{y-y_J(\s)}}{\sqrt{B(\s)}}\right)\,,
\label{vari_profile3}
\end{equation}
with $\tilde L=\frac{4\pi}{3}LT$ and the additional simplification of assuming that the bounce is a straight line in the $y_J,B$ plane, 
\begin{eqnarray}
y_J(\s)&=& y_{J,tv} -  (y_{J,tv}-1)\alpha(\s) \ ,\nonumber\\
B(\s)&=&B_{tv}(1-\alpha(\s)) \ ,
\label{y0Balpha}
\end{eqnarray}
where the $tv$ labels mean ``true vacuum''.
The latter corresponds to $\alpha(\s)=0$, while the false vacuum to $\alpha(\s)=1$.
The variational ansatz (\ref{vari_profile3}) turns out to provide an excellent approximation to the two known solutions with $\alpha(\sigma)=0, 1$.
Plugging (\ref{vari_profile3}) and (\ref{y0Balpha}) into (\ref{vari_acti2}), we can derive the equation of motion for
$\alpha(\s)$ and look for the solution that satisfies  
\be
\alpha'(0)=0 \  , \q \q \q \lim_{\s\to\infty}\alpha(\s) = 1 \ .
\ee

In this way, it is possible to find the following expression for the (rescaled) $O(3)$-symmetric on-shell bounce action, which provides a very good fit of the numerical results,
\begin{equation}
\Delta \tilde S \approx \begin{cases}
0.555 \tilde L^5 \qquad\qquad\qquad\qquad\  \ (\tilde L \leq 0.31)\\
4.61 \cdot 10^{-6} \exp (18.8 \tilde L) \qquad\, ( 0.31\leq \tilde L  \leq 0.57)\\
\frac{0.000467}{(0.6442\, -\tilde L)^2}+\frac{0.00937}{0.6442\, -\tilde L}\qquad \q  \,(\tilde L \geq 0.57)
\end{cases}
\label{fits1An}
\end{equation}
The possible occurrence of $O(4)$ symmetric bounces in the present setup is problematic. The blackening factor $f_T(u)$ in the background (\ref{wittenbackground}) breaks the $O(4)$-symmetry and an ansatz of the form $x_4(u,\rho)$ where $\rho$ is the 4d Euclidean radial coordinate is not consistent with the equations of motion. In \cite{Bigazzi:2020phm}, just as a reference, we have considered a ``naive $O(4)$ configuration" obtained by simply considering the measure $d ^4 x$ to be given by $d \O_3 d \r \r ^3$, where $d \O_3$ is the measure of the three-sphere. We have not found convincing indications that a real $O(4)$ configuration can actually be achieved for the chiral symmetry breaking transition. Thus, in this case, we have decided to focus only on the $O(3)$ symmetric one.

\subsection{GW parameters}
\label{estimates}
It is instructive to give an estimate of how the relevant parameters entering the computation of the stochastic GW spectra depend on the WSS parameters. In this subsection we focus on the confinement/deconfinement phase transition and neglect the flavor contributions. 

In the small temperature regime, when the $O(4)$ symmetric bounce dominates the transition, the bubble nucleation rate (\ref{Gamma2}) can be easily computed using the relations (\ref{fitsO4})  giving
\be
\Gamma(T)= M_{KK}^4 \frac{(S_{4,B})^2}{4\pi^2\bar\rho_w^4} e^{-S_{4,B}} \approx M_{KK}^4 \frac{c_4 ^2 g^2 \bar T^8}{4\pi^2 b^4} e^{- c_4\, g\, \bar T^3}\,,
\ee
where $c_4\approx 0.39$ and $b\approx4$. The rate has a peak at $\bar T= \bar T_m =[8/(3c_4 g)]^{1/3}$ where $\Gamma(T_m)\sim M_{KK}^4 g^{-2/3}$. Hence, increasing $g$, both $\bar T_m$ and $\Gamma(T_m)$ decrease. This qualitative behavior holds in general, beyond the small temperature regime, as can be appreciated by the analysis of the rates done in \cite{Bigazzi:2020phm}. 

The nucleation temperature can be estimated using eq.~(\ref{tildebeta}) with $r=5/3$ and (\ref{conditiontnI}), giving the relation
\be
\frac{\Gamma(T_n)}{ H(T_n)^4}\approx \frac{9}{5} c_4 g {\bar T}_n^3\,,
\label{deftna1}
\ee
where, since we are working in a small temperature regime $\bar T\ll1$, the Hubble parameter is dominated by the vacuum energy contribution
\be
H(T_n)^4 \approx \frac{g^2}{3^{16}\pi^4} M_{KK}^4 \left(\frac{M_{KK}}{M_{pl}}\right)^4\,.
\label{acca4}
\ee
Hence, from (\ref{deftna1}) we get
\be
\label{stepestimate}
\bar T_n^5 e^{-c_4\, g\, \bar T_n^3} \approx \frac{4 b^4 g}{3^{14}5\pi^2 c_4}\left(\frac{M_{KK}}{M_{pl}}\right)^4\ .
\ee 
Now, if, for $\bar T_n\ll1$ and $g\gg1$, we have
\be
c_4\, g\, \bar T_n^3 \gg \frac{5}{3} | \log (c_4 g \bar T_n^3) |\ ,
\ee
i.e. if $c_4 g \bar T_n^3 \gg 1.7$, from (\ref{stepestimate}) we get 
\be
{\bar T}_n^3 \approx \frac{4}{c_4\, g}\log\left(\frac{M_{pl}}{g^{2/3} M_{KK}}\right) + {\cal O}(1/g)\,.
\label{bartnO4}
\ee
The nucleation temperature, in the limit where the above approximations hold, decreases when $g$ and $M_{KK}$ increase and keeps being much smaller than the critical temperature. Supercooling is thus enhanced when $g$ and $M_{KK}$ grow.   
 
In the same limits we can estimate the percolation temperature from eq.~(\ref{percolationtemperatureradiationnew}). If $v\sim1$ we find that
\be
{\bar T}_p^3 \approx {\bar T}_n^3 +{\cal O}(1/g)\,.
\ee

Using the above results and approximations we can also estimate the other relevant parameter, defined in eq.~(\ref{minusfour}), as
 \be
\frac{\beta}{H _*}|_{T_p} = - \frac{3}{5}T_p \frac{d \log\Gamma}{d T}|_{T_p}\approx \frac{3}{5}\left(3 c_4\, g\, \bar T_p^3 -8\right) \approx\frac{3}{5}\left[ 12\log\left(\frac{M_{pl}}{g^{2/3} M_{KK}}\right) -8\right]\,,
 \ee  
up to a velocity dependent term. In the small temperature regime, we thus find that $\beta/H _*$ slightly decreases as $M_{KK}$ and $g$ increase. 

When the $O(3)$ configuration dominates, using the small $\bar T$ expression in the first row of eq. (\ref{fitsO3}), and taking the large $g$ limit, we can analogously get the nucleation and percolation temperatures. In this case the bubble nucleation rate is given by
\be
\Gamma(T) = M_{KK}^4 \frac{{\bar T}^4}{(2\pi)^4} \left(\frac{S_{3,B}}{2\pi T}\right)^{3/2} e^{-S_{3,B}/T}\approx M_{KK}^4\frac{{\bar T}^{31/4}}{(2\pi)^{11/2}} (c_3 g)^{3/2} e^{- c_3 g {\bar T}^{5/2}}\,,
\ee
where $c_3\approx 0.32$. It has a peak at $\bar T_m = [31/(10 c_3 g)]^{2/5}$, where $\Gamma(T_m)\sim M_{KK}^4 g^{-8/5}$. Again, both $\bar T_m$ and $\Gamma(T_m)$ decrease while increasing $g$, in agreement with the more general numerical analysis done in \cite{Bigazzi:2020phm}.

The relation (\ref{conditiontnI}) determining the nucleation temperature now reads
\be
\frac{\Gamma(T_n)}{ H(T_n)^4}\approx \frac{3}{2} c_3 g {\bar T}_n^{5/2}\,,
\label{deftna}
\ee
where, again, the Hubble parameter is approximated by (\ref{acca4}). If, for $\bar T_n\ll1$ and $g\gg1$, we have
\be
c_3\, g\, \bar T_n^{5/2} \gg \frac{21}{10} | \log (c_3 g \bar T_n^{5/2}) |\ ,
\ee
i.e. if $c_3 g \bar T_n^{5/2} \gg 2.1$, we get
\be
\bar T_n^{5/2} \approx \frac{4}{c_3\, g}\log\left(\frac{M_{pl}}{M_{KK} g^{9/10}}\right) + {\cal O}(1/g)\,.
\ee
Again, $\bar T_n$ decreases as $g$ and $M_{KK}$ increase. 

In the same limits ad before, the percolation temperature approximately coincides with the nucleation temperature and
\be
\frac{\beta}{H_*}|_{T_p}\approx \frac{3}{5}\left(\frac{5}{2} c_3 g \bar T_p^{5/2} -\frac{31}{4}\right)\approx  \frac{3}{5}\left[10\log\left(\frac{M_{pl}}{M_{KK} g^{9/10}}\right) -\frac{31}{4}\right]\,,
\label{betasuHthick}
\ee
up to a velocity dependent term.  This parameter decreases as $g$ and $M_{KK}$ increase. 

For strong supercooling, in both the $O(3)$ and the $O(4)$-symmetric cases, the reheating temperature calculated from (\ref{condtR}) reads
\be 
\bar T_{R} \approx \left(\frac{160}{3^6 g^*_{SM}}\right)^{1/4} g^{1/4}\,,
\label{terreapprox}
\ee
where $g^*_{SM}=g^*_{SM}(T_R)={\cal O}(100)$. The reheating temperature is thus independent from $M_{KK}$ and increases with $g$. When $g\gg1$, it is parametrically larger than the critical temperature. 

Finally, using the definition (\ref{alphacap}), it is possible to estimate the parameter $\alpha$, measuring the relative energy released during the transition. In the small temperature regime it reads
\be
\alpha(T_p)\approx \frac{1}{5{\bar T_p}^6}\gg1\,.
\ee

\end{document}